\documentclass[12pt]{article}
\pdfoutput=1

\usepackage{amsmath, amssymb, latexsym}
\usepackage{mathtools}
\usepackage{graphicx}
\usepackage{wrapfig}
\usepackage{xcolor}
\usepackage{mathrsfs}
\input{epsf.sty}
\definecolor{brown}{cmyk}{0,1,.9,.2}

\def\timesbox{\hbox{$\scriptscriptstyle\times$}}
\def\ant{ {{\lower 1ex  \timesbox} \atop {\raise 1.5ex  \timesbox}}}
\newcommand\ZZ{{\hbox{ Z\kern-1.6mm Z}}}
\newcommand{\Iop}{\relax{\rm I\kern-.18em I}}
\newcommand{\Lop}{\relax{\rm I\kern-.18em L}}
\newcommand{\dop}{\relax{\rm I\kern-.8em d}}
\newcommand{\one}{{\hbox{ 1\kern-1.2mm l}}}

\newcommand{\beq}{\begin{equation}}
\newcommand{\eeq}{\end{equation}}
\newcommand{\bea}{\begin{eqnarray}}
\newcommand{\eea}{\end{eqnarray}}
\newcommand{\ra}{\rangle}
\newcommand{\la}{\langle}
\newcommand{\lt}{\left}
\newcommand{\rt}{\right}

\newcommand{\del}{\partial}

\newcommand{\al}{\alpha}

\newcommand{\dlt}{\delta}
\newcommand{\eps}{\epsilon}

\newcommand{\s}{\sigma}

\newcommand{\omg}{\omega}

\newcommand{\Dlt}{\Delta}

\newcommand{\Omg}{\Omega}

\newcommand{\Wb}{\overline{W}}

\newcommand{\gammalb}{{\underline \gamma}}

\newcommand{\Xh}{\hat X}

\newcommand{\xih}{\hat \xi}

\newcommand{\cA}{{\cal A}}

\newcommand{\cC}{{\cal C}}

\newcommand{\cE}{{\cal E}}
\newcommand{\cF}{{\mathcal F}}

\newcommand{\cH}{{\cal H}}

\newcommand{\cL}{{\cal L}}
\newcommand{\cLM}{{\cal LM}}
\newcommand{\cM}{{\mathcal M}}

\newcommand{\cP}{{\cal P}}

\newcommand{\cT}{{\cal T}}

\newcommand{\cV}{{\cal V}}
\newcommand{\cX}{{\cal X}}

\newcommand{\sX}{\mathscr{X}}

\newcommand{\bbR}{\mathbb{R}}

\newcommand{\bbZ}{\mathbb{Z}}

\newcommand{\rmj}{\mathrm{j}}

\begin{document}

{}~
{}~

\vskip 2cm

\centerline{\Large \bf Covariant Closed String Bits - Classical Theory} 
\vspace{.4in}

\medskip

\vspace*{4.0ex}

\centerline{\large \rm Partha Mukhopadhyay }

\vspace*{4.0ex}

\centerline{\large \it The Institute of Mathematical Sciences}
\centerline{\large \it C.I.T. Campus, Taramani}
\centerline{\large \it Chennai 600113, India}
\centerline{\large \it \&}
\centerline{\large \it Homi Bhaba National Institute}
\centerline{\large \it Anushakti Nagar, Mumbai 400094,India}

\medskip

\centerline{E-mail: parthamu@imsc.res.in}

\vspace*{5.0ex}

\centerline{\bf Abstract}
\bigskip

We study lattice wouldsheet theory with continuous time describing free motion of a system of bound string bits. We use a non-local lattice derivative that allows us to preserve all the symmetries of the continuum including the worldsheet local symmetries. There exists a ``local correspondence'' between the continuum and lattice theories in the sense that every local dynamical or constraint equation in the continuum also holds true on the lattice, site-wise. We perform a detailed symmetry analysis for the bits and establish conservation laws. In particular, for a bosonic non-linear sigma model with arbitrary target space, we demonstrate both the global symmetry algebra and classical Virasoro algebra (in position space) on the lattice. Our construction is generalizable to higher dimensions for any generally covariant theory that can be studied by expanding around a globally hyperbolic spacetime with conformally flat Cauchy slices. 

\newpage

\tableofcontents

\baselineskip=18pt

\section{Motivation and summary}
\label{s:motiv}

Usually in lattice field theories Poincar\'e invariance is sacrificed \cite{wilson, kogut}. The finite difference approximation for the derivative does not satisfy the Leibniz rule for a product of functions \cite{spiegel}. The same is also responsible for the fermion doubling problem, the associated loss of chirality \cite{NNtheorem} and challenges in preserving supersymmetry \cite{dondi}. The consequences of Nilsen-Ninomiya theorem \cite{NNtheorem} can be evaded by considering non-local lattice derivatives, e.g. the SLAC derivative \cite{slac}, which enabled attempts to formulate full supersymmetry on a lattice \cite{dondi, bartels, nojiri} (see also \cite{no-go-1, no-go-2}). However, SLAC derivative has been shown \cite{karsten-smit} not to have the desired locality property\footnote{Effectively this renders the lattice theory non-renormalizable due to the requirement of non-local counter terms in the lattice perturbation theory. } in the continuum limit except for perturbation theory for low dimensional Wess-Zumino models \cite{local-cont}. Because of this, though important for preserving spacetime symmetries \cite{no-go-1, no-go-2}, this non-local approach is considered not suitable for gauge theories. 

Lattice techniques have also been used in the context of AdS/CFT correspondence \cite{ads/cft}. See \cite{joseph} for a recent review (of reviews). In particular, its application to the study of semi-classical quantum effects around a classical solution of Green-Schwarz superstring in Type IIB $AdS_5 \times S^5$ background \cite{metsaev} has been considered in \cite{latt-WS, bliard, latt-WS-rev}. Worldsheet theories are different from the other field theories usually considered on a lattice in the sense that they possess local spacetime symmetries. One cannot adopt the Wilsonian approach \cite{wilson} of dealing with local gauge symmetries in this case. Rather, a quantum BRST consistency condition should be used. Therefore, whether a given UV regulator correctly defines the quantum theory around a gauge fixed classical solution or not needs to be carefully investigated \cite{roiban1, roiban2}. On a lattice, one usually ensures recovery of global spacetime symmetries in the continuum limit using the lattice perturbation theory. The results of the latter in turn are crucially used to formulate the non-perturbative computations (see e.g. \cite{capitani} for a review). For theories with local spacetime symmetries, it would be ideal to develop a BRST framework for lattice computations, something that is presently lacking\footnote{In absence of a BRST formulation, perturbation theory was used in \cite{bliard} to investigate the issue in the form of renormalizability and compared with the continuum results.}.  

If the BRST formulation can indeed be done on a lattice, then a more physical way of looking at the lattice worldsheet theory (with continuous time \cite{kogut}) would be to view it as describing the relativistic motion of a system of bound string bits. A note elaborating on this perspective has been presented in Appendix \ref{a:perspective}. A very different approach to string bits has been pursued for long in \cite{giles, bergman, sun}. This is an inherently light-cone approach in flat space and therefore suffers from the same difficulties mentioned above. 

With the above motivation in mind, in this work we shall consider a variant of the non-local derivative discussed in \cite{dondi}.
This derivative involves logarithmic expansions in the difference operators (see eq.(\ref{LDD-def})) and is hereafter referred to as logarithmic discrete derivative (LDD). The operator will be studied to some extent in this work and more rigorously in future works \cite{covlatt2, covlatt3}. These works together will offer a complete proposal for a spacetime-symmetry-preserving lattice construction. 

Like all the existing non-local treatments, LDD is also related to SLAC derivative in a certain sense. However, our approach of how it is used is distinct from all previous works in the literature. A more precise comparison will be discussed in \cite{covlatt2}. 
Below we mention a few crucial aspects of our construction that make it distinct. 
\begin{enumerate}
\item
{\bf Continuation to continuum:} Given a lattice function, there exist infinite number of smooth continuations of it to the continuum. In our construction a unique choice is made. LDD acting on the lattice function gives the same result as the continuum derivative would by acting on the above continuation (see \S \ref{s:ADM} for further discussion on this). The precise choice will be spelled out in \cite{covlatt2}. 

Similar continuation was discussed in \cite{bartels} which was termed as interpolating field. However, our treatment of the lattice derivative is different, as mentioned above. 

\item
{\bf Calculus of LDD:} We shall show in \cite{covlatt3} that the calculus of LDD is identical to that of the continuum derivative (whenever the action of LDD is well defined of course). This, in particular, includes the Leibniz rule and the partial integration law involving arbitrary number of fields (see Appendix \ref{a:leibniz} for further discussion).  

\item
{\bf Infinitesimal symmetries:} Infinitesimal symmetry transformations and Noether's theorem of the continuum hold true in our construction. This is in contrast to the claim in \cite{bartels}. In fact, the purpose of the present work is to present a detailed analysis in position space to exhibit this feature. 

\item
{\bf Local correspondence:} As will be discussed in \S \ref{symmetries}, any local dynamical or constraint equation in the continuum theory also holds true in the lattice theory, site-wise. This plays a crucial role in the analysis of the lattice theory, in particular its symmetries. This also implies that a gauge theory on the lattice can be considered whose Lagrangian is site-wise local gauge invariant. This may help resolving the problem of non-locality of the SLAC derivative mentioned earlier.  

\end{enumerate}

The rest of the paper is organized as follows. In \S \ref{definition} we describe the general approach of constructing a covariant sting bits model of which non-linear sigma model (NLSM) is an example. In \S \ref{laws} we discuss the local correspondence and the Gauss's law on the lattice. Infinitesimal symmetries and conservation laws are discussed in \S \ref{symmetries} where we present explicit results for string bits NLSM both in configuration space and canonical approaches. In particular, we demonstrate both the global symmetry algebra as given by the isometry algebra of the target space and classical Virasoro algebra for the bits. In \S \ref{generalization}  we generalize the construction to higher dimensions and finally conclude in \S \ref{conclusion}. Many technical details are presented in several appendices which we refer to as we go along.

\section{Definition of the Covariant String Bits Model}
\label{definition}

Below we discuss three different ways of describing the covariant string bits model. The first one, given in \S \ref{s:ADM}, is obtained by directly discretizing the continuum theory written in an ADM frame. However, the local diffeomorphism is not manifest in this description. Two ways of describing the model with manifest covariance is described in \S \ref{s:cov}. One of them is a continuum description which uses a filter which  restricts the computation to the lattice. The other one is a purely discrete description for which a 2D tensor calculus on the lattice needs to be defined. The continuum limit of our model is discussed in Appendix \ref{a:cont-lim}. 

\subsection{ADM description}
\label{s:ADM}

Given a worldsheet action, here we spell out the general procedure for constructing the corresponding discrete version. Let $(W, \gamma)$ be the Lorentzian worldsheet, which is topologically a cylinder, with $\gamma_{ab}(\xi)$ the worldsheet metric in a general coordinate system $\xi^a$ ($a = 0, 1$). Then consider an action,
\bea
S &=& \int_W d^2\xi \sqrt{\gamma} \cL(\gamma, Z) ~, \quad \gamma = - \det(\gamma_{ab}) ~,
\label{action-cont}
\eea
where $\cL$ can in general depend on higher derivatives of $\gamma_{ab}$ and $Z^{\al}$ (string coordinates). To define the discrete version, we first perform ADM decomposition of the two dimensional geometry as reviewed in Appendix \ref{a:ADM}. Then the above integral takes the following form,
\bea
S &=& \int_{\bbR \times S^1} (d\tau \beta) (d\s e) \hat \cL(\beta, e, l, \hat Z) ~,
\label{S-ADM}
\eea
where $\xih = (\tau, \s)$ is the coordinate system adopted to the foliation such that $-\infty < \tau < \infty$ is the time coordinate and $\s \in [0,2\pi]$ parametrizes the Cauchy circle. $\beta > 0$ and $e > 0$ are the einbeins on $\bbR$ and $S^1$ respectively (see Appendix \ref{a:ADM} for further details) and,
\bea
\hat \cL(\beta, e, l, \hat Z) &=& \cL(\hat \gamma, \hat Z) ~,
\label{cLhat-cL}
\eea
with,
\bea
\hat \gamma_{ab} &=& \begin{pmatrix} 
-\beta^2 + l^2 & e l \cr e l & e^2 
\end{pmatrix} ~,
\label{gamma-hat-mat}
\eea
where a hat on a tensor field refers to the ADM frame. 

We now discretize the $\s$-coordinate in ADM frame in the following manner while keeping $\tau$ continuous, 
\bea
\s_p = (p-1) h ~, \quad h={2\pi \over N}~, \quad p=1, \cdots , N ~.
\label{sigma-p}
\eea
Given any worldsheet field $\hat X(\tau,\s)$ in ADM frame, we denote the corresponding discrete analogue by $\hat X_p(\tau) :=\hat X(\tau, \s_p)$. The lattice version of (\ref{S-ADM}) is then given by,
\bea
S_{B} &=& h \sum_{p=1}^{N} \int_{\bbR} d\tau  \beta_p e_p \hat \cL(\beta_p, e_p, l_p, \hat Z_p) ~,
\label{discrete-ADM}
\eea
where all the continuous $\s$-derivatives appearing in $\hat \cL$ in (\ref{S-ADM}) are replaced by LDD,
\bea
\del_{\s} &=& {1\over 2} (\del_+ + \del_-) ~,  \cr
\del_{\pm} &:=& \pm {1\over h} \ln (1\pm \Dlt_{\pm}) = \mp {1\over h} \sum_{n=1}^{\infty} {1\over n} (\mp \Dlt_{\pm})^n ~,
\label{LDD-def}
\eea
where the forward and backward difference operators are given by,
\bea
\Dlt_{\pm} f_p = \pm f_{p\pm 1} \mp f_p ~,
\label{difference-op}
\eea
respectively. All the fields appearing in (\ref{discrete-ADM}) satisfy periodic boundary condition: $f_{p+N}=f_p$.

We use the same notation $\del_{\s}$ for the discrete derivative as well as the continuous one to reflect a property to be discussed below. To keep our notations unambiguous we adopt the rule that while acting on a discrete quantity (such as $\hat X_p$) $\del_{\s}$ automatically refers to the discrete derivative. In case of a potential confusion we explicitly use the notation\footnote{When there is no chance of confusion, we shall also use a more simplistic notation $\del$ for LDD e.g. in \S \ref{s:canonical} and Appendices \ref{a:leibniz} and \ref{a:delta-ids}. } $\del_{\s_p}$.

We now explain why we choose LDD as the lattice derivative. $\del_+$ can be motivated in the following manner. Given a smooth function $f(\s)$, we may calculate action of the forward difference operator in the following way,
\bea
\Dlt_+ f(\s_p) &=& \lt[ e^{{1 \over h} \del_{\s}} - 1\rt] f(\s) |_{\s=\s_p} ~,
\eea
where we have first used (\ref{difference-op}) and then Taylor expanded $f(\s_{p+1})$ around $f(\s_p)$. On an infinite lattice, we now formally invert the above relation to obtain,
\bea
\del_{\s} f(\s)|_{\s=\s_p} &=& \del_+ f_p ~, \quad f_p := f (\s_p) .
\label{der-consistency0}
\eea
While on the left hand side we have a derivative in the continuum, the right hand side is computed purely from the knowledge of $\{f_p, p \in \bbZ \}$. One can make similar observation for the backward difference operator as well\footnote{In the theory of finite difference calculus \cite{spiegel}, $\del_+$ is considered to be a formal expression where a truncation of the logarithmic expansion provides only an approximation for the continuous derivative. Here we consider the following view. Given all possible lattice field configurations $\sX=\{ \{f_q \}, ~q=1, 2, \cdots , N \}$, the complete operator in (\ref{LDD-def}) (without any approximation) can be viewed as a genuine lattice operator satisfying all the crucial properties of the usual differential calculus, thanks to (\ref{der-consistency}), provided its action on $\sX$ is well defined. \label{LDD-exp}}. Therefore, we may conclude that given any smooth function $f(\s)$, the following is true,
\bea
\del_{\s} f(\s)|_{\s=\s_p} &=& \del_{\s} f_p ~, 
\label{der-consistency}
\eea
provided the right hand side is well defined. This argument can be extended to all Fourier transformable functions, either on $\bbR$ or on a line segment, in which case the above equation has to hold true only for the Fourier modes. In our approach to construct a covariant theory of string bits, eq.(\ref{der-consistency}) will be crucially used. The latter will be investigated more elaborately in a future work \cite{covlatt3}. We give preliminary evidence of how the Leibniz rule and partial integration law hold for this operator in Appendix \ref{a:leibniz}. 

Note that we have considered a uniform lattice in a specific frame (ADM) where $\s$ does not measure invariant length and LDD satisfies,
\bea
\del_{\s} \s_p = 1 ~,
\eea
which is a fundamental requirement. It will be generalized to other frames in the next subsection. The specific linear combination of $\del_{\pm}$ considered in (\ref{LDD-def}) allows one to satisfy (as proved in Appendix \ref{a:delta-ids}) a specific Kronecker delta identity, namely (\ref{der-kronecker-1}). Such identities are the discrete analogues of the corresponding ones satisfied by the Dirac delta function in the continuum and are required to write the constraint algebra (\ref{witt-alg}) of string bits NLSM in the desired form. 

We now clarify an important subtlety in the previous discussion. In eq.(\ref{der-consistency}), $f(\s)$ is given and $f_p$ is obtained from that. But in a lattice field theory one starts with a lattice field configuration $\{f_p\}$. Therefore, one has to find a smooth function $f(\s)$ which satisfies,
\bea
f(\s_p) = f_p ~, \quad \forall p = 1, \cdots, N~.
\label{continuation}
\eea
Such a function is not unique as the above equation is invariant under,
\bea
f(\s) \to f(\s) + C(\s) ~,
\eea
where $C(\s)$ is smooth but constant on the lattice: $C(\s_p) = C(\s_p+h) = 0$. Therefore, the result of the left had side of (\ref{der-consistency}) is ambiguous. Given $\{f_p\}$, the same is true for the right hand side as well. This is because although $\Dlt_{\pm} C_p = 0$, $\del_{\pm} \sim \pm {1\over h} \ln 1$ suffers from logarithmic branch cut ambiguity. Therefore, each time LDD operates on a lattice field, this ambiguity needs to be fixed suitably in order for LDD to satisfy the same differential calculus as the continuum derivative\footnote{$C(\s)$ is periodic and is Fourier transformable. Therefore, the ambiguity needs to be fixed for each Fourier mode. }. This procedure also uniquely fixes the continuum function $f(\s)$ in (\ref{continuation}). A prescription for fixing the continuation-ambiguity in this way will be suggested in \cite{covlatt2}. In this work, whenever we use (\ref{der-consistency}), we assume that this has already been done consistently.

\subsection{Covariant descriptions}
\label{s:cov}

Given an action in the continuum, equation (\ref{discrete-ADM}), which we refer to as ADM description, provides the fundamental definition of the corresponding discrete model. However, the spacetime symmetries of the continuum are still obscure in this description. We shall now describe a covariant form where the symmetries are manifestly present. In this form, the same lattice expression in (\ref{discrete-ADM}) is given by,  
\bea
S_B &=& \int_{\Wb} d^2 \xi \sqrt{\gamma}  \cL(\gamma, Z ) \cF (\xi, \{ \xi_p \}) ~,  
\label{continuous-cov}
\eea
where $\Wb$ is an auxiliary worldsheet with the same topology of $W$ and,
\bea
\cF ( \xi, \{ \xi_p \} ) &=& h \sum_{p=1}^{N} \int_{\bbR} d\tau e_p \beta_p \dlt^2(\xi, \xi_p)  ~,
\label{filter}
\eea
is a filter which reduces the two dimensional integral to ones along the bits worldlines $\{\xi_p\}$ and,
\bea
\dlt^2(\xi, \xi_p) &=& {\dlt^2(\xi-\xi_p) \over (\gamma(\xi))^{1/4} (\gamma(\xi_p))^{1/4} } ~,
\label{dlt2}
\eea
is the bi-scalar Dirac delta distribution. $\xi_p$ is the worldline for the $p$-th string bit in general coordinates $\xi$. In general $\xi_p$ depends on both $\tau$ and $\s_p$, as it is obtained by performing the necessary general coordinate transformation (GCT) $\xih \to \xi$ and restricting that to the $p$-th worldline in ADM frame, 
\bea
\xi_p := \xi (\xih_p) ~, \quad \lt[ \xih_p = (\tau, \s_p) \rt] ~.
\label{bits-GC-1}
\eea
Notice that the RHS suffers from the same ambiguity in choosing the continuum function $\xi(\hat \xi)$ as mentioned before. We assume that this has been fixed and we shall continue to do so for all cases where continuum functions appear in lattice equations. In the context of the aforementioned GCT, such a fixation has an impact on the computation of the volume of transformation, something that we shall not discuss in this work. We however make further comments on this in point $2$ below. 

We now discuss a number of properties of the filter and their consequences. 
\begin{enumerate} 
\item
{\bf Multi-point scalar field under GCT:} By construction, the filter is a multi-point scalar under a GCT $\xi \to \xi'$,
\bea
\cF(\xi, \{\xi_p\} ) \to \cF'(\xi', \{\xi'_p\}) = \cF(\xi, \{\xi_p\}) ~.
\eea
This results in the invariance of $S_B$ in (\ref{continuous-cov}) under the same. Note that $\xih^a_p=(\tau, \s_p)$, $e(\xih)$ and $\beta(\xih)$ appearing in (\ref{filter}) do not transform under such a transformation, simply because they refer to a fixed ADM frame. By performing the GCT $\xi \to \xih^a = (\tau, \s)$ in (\ref{continuous-cov}) it is straightforward to establish the equivalence between the ADM and covariant description, namely, 
\bea
h \sum_{p=1}^{N} \int_{\bbR} d\tau e_p(\tau) \beta_p(\tau) \cL(\hat \gamma_p, \hat Z_p) &=& \int_{\Wb} d^2 \xi \sqrt{\gamma (\xi) } \cL(\gamma (\xi), Z(\xi) ) \cF(\xi , \{\xi_p\}) ~. \cr
&&
\label{ADM-continuous-eq}
\eea

\item
{\bf Weyl invariance:} The filter is Weyl invariant as the ADM frame does transform under Weyl transformation,
\bea
\gamma_{ab} \to \Omg \gamma_{ab}~, \quad \hat \gamma_{ab} \to \hat \Omg \hat \gamma_{ab} ~, \quad e_p \to \sqrt{\hat \Omg_p} e_p ~, \quad \beta_p \to \sqrt{\hat \Omg_p} \beta_p ~,
\eea
where $\Omg(\xi) = \hat \Omg(\hat \xi)$. This implies, in particular, that string bits NLSM with an arbitrary target space $(\cM, G)$, 
\bea
\cL(\gamma, Z) &=& -{1\over 4\pi \al'} \gamma^{ab} \del_a Z^{\al} \del_b Z^{\beta} G_{\al \beta}(Z)~,
\label{nlsm-lag}
\eea
admits a remnant of the two-dimensional (diff $\times$ Weyl) local symmetry of the original continuum theory. Here, by ``remnant'' we simply refer to the fact that the ``volume'' of symmetries is reduced from that of the continuum. The generators of the local symmetries of the bits model are obtained by restricting the generators of the same for the continuum theory on the bit-worldlines and removing the continuation-ambiguity as discussed at the end of \S \ref{s:ADM}

\item
{\bf Field independence:} The filter is invariant under the variations of the fields. By recognizing the following form of the filter,
\bea
\cF (\xi, \{\xi_p\}) &=& h\sum_{p=1}^N \int_{\bbR} d\tau \sqrt{\hat \gamma (\xih_p)} \dlt^2 (\xi, \xi_p)~, 
\label{filter-cov}
\eea
it is straightforward to establish that,
\bea
\dlt \cF(\xi, \{\xi_p\}) &=& h\sum_{p=1}^N \int_{\bbR} d\tau \sqrt{\hat \gamma (\xih_p) } \cr
&& \lt[{1\over 2} \hat \gamma^{ab}(\xih_p) \dlt \hat \gamma_{ab}(\xih_p) - {1\over 4} \gamma^{ab}(\xi) \dlt \gamma_{ab}(\xi) - {1\over 4} \gamma^{ab}(\xi_p) \dlt \gamma_{ab}(\xi_p) \rt] \dlt^2 (\xi, \xi_p) ~, \cr
&=& 0~.
\eea
Since each term within the square bracket is a scalar, it can be evaluated at any particular frame, e.g. ADM frame. Then the sum vanishes as $\xi \to \xi_p$, which is imposed by the delta function. 

\item
{\bf Covariant constancy:} We prove in Appendix \ref{a:delta-ids} that the filter is covariantly constant in the following sense,
\bea
\int d^2 \xi D_a \cF(\xi, \{\xi_p\}) X^a(\xi) &=& 0 ~, \quad \forall X ~, \hbox{such that } \lim_{\tau \to \pm \infty} X_p^a = 0 ~, \forall p=1, \cdots, N~.
\cr &&
\eea
where $D_a$ is the metric compatible covariant derivative on the worldsheet.

\end{enumerate}
The local correspondence between continuum and lattice theories mentioned earlier, which we discuss in detail in \S \ref{laws}, follows from the field independence and covariant constancy of the filter. 

Now notice that in deriving (\ref{ADM-continuous-eq}) one first exploits general covariance to go to the ADM frame on the right hand side and then integrates the filter out to arrive at the left hand side, implicitly assuming the property (\ref{der-consistency}). However, the filter could be integrated out in a general coordinate system itself without having to go to the ADM frame first. In that case, one arrives at the following ``discrete covariant model'', 
\bea
S_B &=& h \sum_{p=1}^{N} \int_{\bbR} d\tau e_p \beta_p \cL (\gamma_p, Z_p) ~,
\label{discrete-cov}
\eea
where $\gamma_{p ab}(\tau) := \gamma_{ab}(\xi_p)$
and so on. 

We now explain in what sense the above is a discrete covariant model. The action (\ref{discrete-cov}) may be considered as a lattice model in its own right without referring to any auxiliary worldsheet. In order to do so discrete version of the transformation rules under finite GCT needs to be formulated as it is done below. 

The discrete space under consideration, 
\bea
B_1^2(N) &=& \{ (\tau, \s_p) | \tau \in [\tau_1, \tau_2], p=1, \cdots , N \} ~,
\label{B-space}
\eea
with boundary,
\bea
\del B_1^2 = \del_1 B \cup \del_2 B ~, \quad \del_i B \equiv \{ (\tau_i, \s_p) | p=1,\cdots ,N  \} ~, ~ i=1, 2 ~,
\eea
is a finite world-tube of the string bits\footnote{Note that $B_1^2(N)$ is a closed subset of the auxiliary worldsheet with boundary $\overline{W}^2_1$ (see eq.(\ref{del-Wb})), the latter being homeomorphic to $[\tau_1, \tau_2] \times S^1$, 
\bea
B^2_1(N) \subset \overline{W}^2_1 ~.
\eea
\label{B-Wbar}}. We may identify, 
\bea
\hat \varphi (w_p) = \xih_p = (\tau, \s_p) ~, \quad \forall w_p \in B_1^2 (N) ~,
\label{globalCS}
\eea
as the ``global coordinate system'', where $\hat \varphi$ is the coordinate map. The derivatives $\{ \del_{\tau}, \del_{\s_p} \}$ span a two-dimensional tangent space $T_{w_p}B^2_1$ associated to $w_p \in B^2_1$. Correspondingly, we define the cotangent space $T^*_{w_p}B^2_1$, dual to $T_{w_p}B^2_1$, which is spanned by the dual basis $\{d\tau, d\s_p\}$ satisfying the following inner product,
\bea
\large \la d\xih_p^a, {\del \over \del \xih_p^b} \large \ra = \dlt^a_b ~.
\label{inner-prod-global}
\eea

A set of general coordinates $\xi_p$ is given by,
\bea
\xi_p^a = \xi^a(\xih_p) ~,
\label{bits-GC-2}
\eea
where $\xi^a$ are lattice functions which are smooth with respect to LDD. After fixing the continuation-ambiguity ambiguity, the map becomes invertible,
\bea
\xih_p^a = \xi^{-1 a} (\xi_p) ~, 
\eea
where $\xi^{-1}$ is inverse of $\xi$. This allows us to define the Jacobian matrix for the transformation and its inverse,
\bea
J_p^a{}_b \equiv \lt( {\del \xi_p^a\over \del \xih_p^b} \rt) ~, \quad (J^{-1}_p)^a{}_b \equiv \lt( {\del \xih_p^a\over \del \xi_p^b} \rt) ~,
\eea
respectively. Given two sets of general coordinates, $\xi^{(1) a}_p$ and $\xi^{(2) a}_p$ and the corresponding Jacobian matrices $J^{(1) a}_p{}_b$ and $J^{(2) a}_p{}_b$ respectively, the transformation $\xi^{(12)}$ directly relating them,
\bea
\xi^{(1)a}_p = \xi^{(12)}(\xi^{(2)}_p) ~,
\eea
admits the corresponding Jacobian matrix $J^{(12)}_p$ which is expressible in terms of $J^{(1) a}_p$ and $J^{(2) a}_p$,
\bea
J^{(12) a}_p{}_b &=& \lt( {\del \xi_p^{(1) a} \over \del \xi_p^{(2) b} } \rt) = J^{(1) a}_p{}_c J^{(2) -1 c}_p{}_b ~.
\eea

Given the above machinery, we are now able to describe the elements of tangent and cotangent spaces in a general coordinate system. The basis elements are given by $\{{\del \over \del \xi_p^a} \}$ and $\{d\xi_p^a\}$ respectively which are related to the ones in the global system (\ref{globalCS}) in the usual manner through the Jacobian factors defined above. As a result, the inner product (\ref{inner-prod-global}) takes the following form in general coordinate system,
\bea
\large \la d\xi_p^a, {\del \over \del \xi_p^b} \large \ra &=& \dlt^a_b ~.
\label{inner-prod-GCS}
\eea
Having thus completed the basic set-up, one is now able to define tensor fields on the lattice exactly in the same manner as it is done in the continuum. 

\section{A Local Correspondence and Gauss's Law}
\label{laws}

Here we discuss two crucial properties of the lattice model which have important consequences for the symmetry analysis to be performed in the next section. The first property is a ``local correspondence'',
\bea
&& \hbox{\it Given a local dynamical or constraint equation in the continuum, the same also holds} \cr
&& \hbox{\it  true on the lattice site-wise, with the continuum $\s$-derivative replaced by LDD.} 
\label{local-correspondence}
\eea
Given that such local equations are typically obtained by taking field variations of the action, here we analyze the correspondence in the context of the latter. In the continuum, if variation of the action with respect to any specific field $X$ is given by,
\bea
\dlt_X S &=& \int_W d^2 \xi \sqrt{\gamma(\xi)} \dlt X(\xi) \cE (\cX(\xi) ) ~,
\label{action-var-cont}
\eea
where $\cE$ is, in general, constructed out of the set $\cX$ of all possible fields and their derivatives in the theory, then the corresponding lattice analogue is given by,
\bea
\dlt_X S_B &=& \int_{\Wb} d^2 \xi \sqrt{\gamma(\xi)} \dlt X(\xi) \cE (\cX(\xi)) \cF (\xi, \{\xi_p \}) ~, \cr
&=& h \sum_{p=1}^N \int_{\bbR} d\tau \sqrt{\hat \gamma_p(\tau)} \dlt X_p(\tau) \cE(\cX_p(\tau)) ~,
\label{action-var-bits}
\eea
where $\cX_p(\tau) = \cX(\xi_p)$ as usual. The above result can be obtained by using both the continuum and discrete covariant descriptions of \S \ref{s:cov} where the definitions of the fundamental functional derivative in the two cases are taken to be,
\bea
{\dlt X(\xi') \over \dlt X(\xi)} = \dlt^2(\xi', \xi) ~, \quad {\dlt X_{p'}(\tau') \over \dlt X_p(\tau)} = {1\over h} \dlt^2_{(p',p)}(\tau',\tau) ~,
\label{fund-func-derv}
\eea
respectively. The lattice version of the bi-scalar delta function is given by,
\bea
\dlt^2_{(p',p)}(\tau',\tau) = {\dlt_{p',p} \dlt(\tau'-\tau) \over \sqrt{e_{p'}(\tau') \beta_{p'}(\tau') e_p(\tau) \beta_p(\tau)} } ~.
\eea
We prove the above result in Appendix \ref{a:technical} where we also show that the two ways of doing the calculation are consistent with each other thanks to a specific identity (\ref{consistency}) involving both Dirac and Kronecker delta functions. 

The second property is the Gauss's theorem. In the discrete language, it is given by,
\bea
\sum_{p=1}^N \int_{\tau_1}^{\tau_2} d\tau \sqrt{\hat \gamma_p(\tau)} D_a J_p^a &=& \sum_{p=1}^N \lt[ \sqrt{\hat \gamma_p(\tau)} \hat J^0_p(\tau) \rt]^{\tau_2}_{\tau_1} ~,
\label{gauss-discrete}
\eea
where $J^a$ is a 2D current density. The same result can also be written, in continuous covariant form, as,
\bea
\int_{\Wb^2_1} d^2\xi \sqrt{\gamma(\xi)} D_a J^a (\xi) \cF(\xi, \{\xi_p\}) &=& \int_{\del \Wb^2_1} dx \sqrt{\gammalb (x)} n_a(x) J^a (\xi) \cF(\xi(x), \{\xi_p\}) ~, 
\label{gauss-cont-cov}
\eea
where $\Wb^2_1$ is a finite auxiliary worldsheet with boundary,
\bea
\del \Wb^2_1 = \del_1 \Wb \cup \del_2 \Wb ~,
\label{del-Wb}
\eea
is the union of two disconnected Cauchy circles $C_i = \del_i \Wb$, $i=1, 2$, in counter clockwise sense when viewed from past. Furthermore, $x$ is a general coordinate on $\del \Wb_1^2$ with $\gammalb$ being the induced metric on it and $n^a$ is the outward unit timelike normal. Therefore, if $C_1$ occurs in the past with respect to $C_2$, then $n^a (C_1)$ and $n^a (C_2)$ are past and future directed respectively. At any point if $n^a$ is future directed, then in the ADM frame it satisfies: $\hat n_{\tau} >0$. This makes sure that the line integral along constant $\s$ in the direction of increasing $\tau$ is positive. Further technicalities associated to the above result are discussed in Appendix \ref{a:technical}. 

\section{Infinitesimal Symmetries and Conservation Laws}
\label{symmetries}

Here we discuss the infinitesimal symmetries of the string bits NLSM and the associated conservation laws. We follow the general framework of \cite{holten}. The configuration space approach is discussed in \S \ref{s:config} where the results are fully covariant. The canonical approach is discussed in \S \ref{s:canonical} where we work in the reduced phase space framework and demonstrate the classical Virasoro algebra of the bits model in position space. 

\subsection{Configuration space approach}
\label{s:config}

In Appendix \ref{a:symmetry}, we generalize the treatment of \cite{holten} to a generally covariant theory and apply that to string bits NLSM corresponding to the Lagrangian density (\ref{nlsm-lag}). The model has the following global symmetries,
\bea
\dlt \gamma_{p ab} = 0 ~, \quad \dlt Z^{\mu}_p = \sum_i \eta_i V_i^{\mu}(Z_p) ~, 
\label{nlsm-global}
\eea
where $\eta_i$'s are infinitesimal constants and $V_i^{\mu}$'s are isometries of the target space $(\cM, G)$ satisfying Killing equations,
\bea
\nabla_{\mu} V_{i\nu} + \nabla_{\nu} V_{i\mu} = 0 ~,
\label{killing}
\eea
where $\nabla_{\mu}$ is the metric compatible covariant derivative in $(\cM, G)$. The corresponding charge in the ADM frame adapted to a specific Cauchy slicing,
\bea
Q(\tau) &=& -a \sum_{i, p} \eta_i  \sqrt{\hat \gamma_p (\tau)} \hat \gamma_p^{0b}(\tau) G_{\mu \nu}(\hat Z_p(\tau)) \del_b \hat Z_p^{\nu}(\tau) V_i^{\mu}(\hat Z_p(\tau)) ~, \quad a = {h\over 2\pi \al'} ~,
\label{global-charge}
\eea
is conserved on-shell. For two variations (\ref{nlsm-global}) given by the sets of parameters $\{\eta^{(1)}_i\}$ and $\{\eta^{(2)}_i\}$, the closed commutator algebra is given by,
\bea
[\dlt^{(1)}, \dlt^{(2)}] Z_p^{\mu} = \dlt^{(3)} Z_p^{\mu} = \sum_k \eta^{(3)}_k V_k^{\mu}(Z_p) ~, &&
\eta^{(3)}_k =  \sum_{i, j} f_{ijk} \eta^{(1)}_i \eta^{(2)}_j ~,
\label{global-alg}
\eea
$f_{ijk}$ being the structure constants of the isometry algebra of target space.

The local symmetries, i.e. Weyl and 2D diffeomorphism, are given by,
\bea
\dlt \gamma_{p ab} = \omg_p \gamma_{p ab} + D_a \kappa_{p b} + D_b \kappa_{p a} ~, \quad
\dlt Z_p^{\mu} = \kappa_p^a \del_a Z_p^{\mu} ~.
\label{nlsm-local}
\eea
The associated constraint 
\bea
G(\tau) = -\sum_p \sqrt{\hat \gamma_p} \hat \kappa_{pb} \hat{T}_p^{0b}  ~, 
\label{local-charge}
\eea
vanishes on-shell, where the stress tensor is given by,
\bea
T_{p ab} &=& a G_{\mu \nu}(Z_p) \lt[ \del_a Z_p^{\mu} \del_b Z_p^{\nu} - {1\over 2} \gamma_{p ab} \gamma_p^{cd} \del_c Z_p^{\mu} \del_d Z_p^{\nu} \rt]  ~.
\label{stress-tensor}
\eea
This is equivalent to vanishing of all components of the stress tensor on-shell,
\bea
T^{ab}_p &=& 0 ~.
\label{constraints}
\eea
Calculation of the commutator of transformations is long and tedious\footnote{The author thanks Chandrima Paul for checking these results.}. The result is given by,
\bea
[\dlt^{(1)}, \dlt^{(2)}] \gamma_{p ab} &=& \dlt^{(3)} \gamma_{p ab} = \omg_p^{(3)} \gamma_{p ab} + D_a \kappa^{(3)}_{p b} + D_b \kappa^{(3)}_{p a}  ~, \cr
[\dlt^{(1)}, \dlt^{(2)}] Z_p^{\mu} &=& \dlt^{(3)} Z_p^{\mu} = \kappa_p^{(3) a} \del_a Z_p^{\mu} ~,  
\label{local-alg-1}
\eea
with,
\bea
\omg_p^{(3)} = 0 ~, \quad \kappa_p^{(3) a} = \kappa_p^{(1) b} D_b \kappa_p^{(2) a} - \kappa_p^{(2) b} D_b \kappa_p^{(1) a} + \kappa_p^{(1) a} \omg_p^{(2)} - \kappa_p^{(2) a} \omg_p^{(1)} ~.
\label{local-alg-2}
\eea

\subsection{Canonical approach}
\label{s:canonical}

In the canonical approach we work with the reduced phase space by fixing the diffeomorphism to unit gauge, 
\bea
\beta_p = e_p = 1 ~, \quad l_p = 0 ~, \quad \forall p =1, \cdots , N ~.
\label{unit-gauge}
\eea
Although we are working in a specific ADM frame, in this part of our discussion we remove hats from the symbols representing dynamical variables and use $\del$ for LDD to reduce clutter. 

The phase space is spanned by $\{(Z_p^{\mu}, P_{p \nu}) \}$, with fundamental Poisson brackets,
\bea
\{Z_p^{\mu}, P_{q \nu} \} = \dlt_{p, q} \dlt^{\mu}{}_{\nu} ~,
\eea
where the conjugate momentum is: $P_{p \mu} = a G_{\mu \nu}(Z_p) \del_{\tau} Z_p^{\nu}$. The global charge is given by,
\bea
Q &=&  \sum_{i, p} \eta_i P_{p \mu} V_i^{\mu}(Z_p) ~.
\eea
This leads to the following Poisson bracket algebra,
\bea
\{ Q^{(1)}, Q^{(2)} \} &=& Q^{(3)} ~,
\eea
with $\eta^{(3)}$ given by (\ref{global-alg}). Similarly, the diffeomorphism constraint,
\bea
G &=& \sum_p [ \kappa_p^+ T_{p+} + \kappa_p^- T_{p-} ] 
\eea
with $\kappa_p^{\pm} = \kappa_p^0 \pm \kappa_p^1$ and $T_{p \pm} = {1\over 2} (T_{p 00} \pm T_{p 01})$ satisfies,
\bea
\{G^{(1)}, G^{(2)} \} = G^{(3)} ~,
\label{PB-alg-local-0}
\eea
where,
\bea
\kappa_p^{(3)+} = \kappa_p^{(1)+} \del \kappa_p^{(2)+} - \kappa_p^{(2)+} \del \kappa_p^{(1)+}  ~, &&
\kappa_p^{(3)-} = -\kappa_p^{(1)-} \del \kappa_p^{(2)-} + \kappa_p^{(2)-} \del \kappa_p^{(1)-}  ~.
\eea
These results can also be written as,
\bea
\kappa_p^{(3)+} = \kappa_p^{(1)+} \del_+ \kappa_p^{(2)+} - \kappa_p^{(2)+}  \del_+ \kappa_p^{(1)+}  ~, &&
\kappa_p^{(3)-} = \kappa_p^{(1)-} \del_- \kappa_p^{(2)-} - \kappa_p^{(2)-} \del_- \kappa_p^{(1)-}  ~,
\label{PB-alg-local}
\eea
where $\del_{\pm} = {1\over 2} ( \del_{\tau} \pm \del)$ if the following conditions are satisfied,
\bea
\del_{\mp} \kappa_p^{\pm} = 0   ~.
\label{local-restrict}
\eea
The results (\ref{PB-alg-local}) are exactly same as the fully covariant result (\ref{local-alg-2}) obtained using configuration space approach with the restriction (\ref{local-restrict}) and $\omg =0 $. These restrictions are required for preserving the unit gauge (\ref{unit-gauge}) chosen a-priori for the present canonical formalism. Another form in which (\ref{PB-alg-local-0}, \ref{PB-alg-local}) can be expressed is the classical Virasoro algebra (in position space),
\bea
\{T_{p \pm}, T_{p' \pm} \} &=& \pm 2 T_{p \pm} \del \dlt_{p, p'} \pm \del T_{p \pm} \dlt_{p, p'}  ~, \cr
\{T_{p +}, T_{p' -} \} &=& 0 ~.
\label{witt-alg}
\eea

Furthermore, Poisson bracket of the global charge with the Hamiltonian: $H = \displaystyle \sum_p T_{p00}$ vanishes,
\bea
\{Q, H\} = 0 ~, 
\eea
thanks to the Killing equation (\ref{killing}). The same for the local constraint vanishes weakly,
\bea
\{G, H\} &=& \sum_p [\kappa_p^+ \del T_{p+} - \kappa_p^- \del T_{p-} ] ~, \cr
&=& \sum_p [\kappa_p^+ \del_+ T_{p+} + \kappa_p^- \del_- T_{p-} ] \approx 0 ~,
\eea
where in the second step we have used,
\bea
\del_{\pm} T_{p \mp} = 0 ~, 
\label{EOM-T}
\eea
which follow from Hamilton's equations of motion.

Calculations leading to the above results are standard in the continuum. For the bits, it goes exactly in the same fashion with the use of Leibniz rule and partial integration laws as discussed in Appendix \ref{a:leibniz}. We end this section by pointing out a few subtleties. First, given any tensor $\cT^{\cdots}_{p \cdots} \equiv \cT^{\cdots}_{\cdots}(Z_p)$ in the target space, we must have,
\bea
\{ \cT^{\cdots}_{p \cdots}, P_{q, \nu} \} &=& \nabla_{\nu} \cT^{\cdots}_{p \cdots} \dlt_{p, q} ~,
\eea
in order to be consistent with general covariance of the target space. This result can also be argued by first going to a suitable Riemann normal coordinate system \cite{eisenhart}, calculate the bracket and then come back to a general coordinate system. This means that the metric is annihilated by the Poisson bracket with $P$. Next, a couple of Kronecker delta function identities involving LDD
which we use (especially in deriving (\ref{witt-alg})), are as follows,
\bea
\del' \dlt_{p, p'} = - \del \dlt_{p, p'} ~,
\label{der-kronecker-1} \\
f_{p'} \del \dlt_{p, p'} = f_p \del \dlt_{p, p'} + \del f_p \dlt_{p, p'} ~.
\label{der-kronecker-2}
\eea
These are discrete analog of the standard continuum identities involving derivative of Dirac delta function. The proofs of the above identities are discussed in Appendix \ref{a:delta-ids}.

\section{Higher Dimensional Generalization}
\label{generalization}

The lattice construction described in \S \ref{definition} is readily generalizable to any generally covariant theory that can be studied by expanding around a globally hyperbolic spacetime with conformally flat Cauchy slices. After performing such an expansion, the action can be written as,
\bea
S &=& \int d^{D+1} \xi \sqrt{\gamma (\xi)} \cL(\gamma, \cX)  ~,
\eea
where $\cX(\xi)$ is the collection of all possible tensor fields other than the background metric $\gamma_{ab}$ and their covariant derivatives (corresponding to $\gamma_{ab}$) that $\cL$ may depend on. $\cL$ can also depend on the curvature tensors of $\gamma_{ab}$ and their higher covariant derivatives which is indicated by its dependence through $\gamma$. The background metric is diffeomorphic to the following ADM form,
\bea
\xi^a &\to & \hat \xi^a = (\tau, \s^i) ~, \quad i =1, \cdots, D ~, \cr
\gamma_{ab} (\xi) &\to & \hat \gamma_{ab}(\hat \xi) = 
\begin{pmatrix}
-\beta^2 + e_{ij} l^i l^j & e_{ij} l^j \cr
e_{ij} l^j & e_{ij} 
\end{pmatrix} ~, \quad e_{ij} = e^2 \dlt_{ij} ~,
\eea
where the lapse function $\beta$, the shift vector $l^i$ and the conformal factor $e^2$ are functions of the ADM-isothermal system $\hat \xi$. 
The corresponding lattice theory is given by (in the three different descriptions described in \S \ref{definition}),
\bea
S_L &=& h \sum_{\vec p} \int d\tau \beta_{\vec p} e_{\vec p}^D \hat \cL ( \beta_{\vec p}, e_{\vec p}, \vec l_{\vec p}, \hat \cX_{\vec p} ) ~, \quad [\hbox{ADM-isothermal}] ~, \cr
&=& \int d^{D+1} \xi \sqrt{\gamma (\xi)} \cL(\gamma, \cX) \cF (\xi, \xi_{\vec p} ) ~, \quad [\hbox{Continuous covariant}] ~, \cr
&=& h \sum_{\vec p} \int d\tau \sqrt{\hat \gamma_{\vec p}} \cL ( \gamma_{\vec p}, \cX_{\vec p} )~, \quad [\hbox{Discrete covariant}] ~,
\label{latt-action-HD}
\eea
where, 
\bea
\hat \cL ( \beta, e, \vec l, \hat \cX ) = \cL (\hat \gamma, \hat \cX) ~.
\eea
The lattice is constructed by discretizing the spatial coordinates $\vec \s = \{ \s^i \}$ in the ADM-isothermal system in the following manner, 
\bea
\vec \s_{\vec p} = \{ \s^i_{p^i} = (p^i-1) a_i \} ~, \quad \vec p = \{ p^i \} ~, \quad p^i = 1, \cdots, N_i ~, \quad a_i = {2\pi \over N_i} ~,
\eea
and $h= \lt(\prod_i a_i\rt)$. As in the string bits case, the lattice fields are identified as $\gamma_{\vec p ab}(\tau) = \gamma_{ab}(\xi_{\vec p})$ and so on with periodic boundary condition, where $\xi_{\vec p}$ is obtained by performing a coordinate transformation on $\hat \xi_{\vec p} = (\tau, \vec \s_{\vec p})$. The discrete tensor calculus, the expression for the filter and the Gauss's law as discussed in the context of sting bits have straightforward generalization to the present higher dimensional case. All the properties of the filter discussed in \S \ref{definition} and the local correspondence discussed in \S \ref{laws} also hold true for the lattice theory given by (\ref{latt-action-HD}).

\section{Conclusion}
\label{conclusion}

In this work we have studied classical lattice worldsheet theory with continuous time where a non-local lattice derivative (LDD) has been used that enables us to preserve all the symmetries of the continuum, including the local worldsheet symmetries. This allows one to study such theories within a BRST framework. The construction is generalizable to higher dimensions for generally covariant theories around any globally hyperbolic spacetime with conformally flat Cauchy slices. 

Such a construction is possible because LDD, once dealt with properly, behaves exactly like the continuum derivative. How exactly to deal with it will be fully clarified in future works \cite{covlatt2, covlatt3}. The goal of the present work was to clarify the functioning of LDD at an intuitive level and  by assuming its actual properties, to develop the tools for covariant analysis of the lattice model in position space. This analysis, in addition to establishing infinitesimal symmetries and the associated conservation laws on the lattice, gives rise to the local correspondence between continuum and lattice theories (see \S \ref{laws}). The latter can have implications in the context of the non-locality problems observed in lattice perturbation theory using SLAC derivative \cite{karsten-smit}.

\appendix

\section{Lattice NLSM as string bits NLSM}
\label{a:perspective}

For string propagation in a general background, the BRST consistency condition gives rise to the equation of motion for the background, e.g. Einstein equation for the target manifold $\cM$. Originally this result was shown in a semi-classical quantization around a classical solution of the corresponding 2D NLSM\footnote{Some of the original references are \cite{friedan, alvarez-gaume, callan}. See also references in \cite{callan} for earlier works.}. The semi-classical expansion itself is formulated \cite{alvarez-gaume} by using the geometric Riemann normal expansion \cite{eisenhart} around a point in $\cM$. There must exist another semi-classical expansion of NLSM where the classical limit is point-like, as is the case for an exactly solvable conformal background like flat space. The aforementioned formulation using Riemann normal expansion is not suitable in this case as the corresponding classical solution is singular. It was suggested in \cite{lsqm} that for such a {\it small string} quantization a loop space framework\footnote{Loop space $\cLM = C^{\infty}(S^1, \cM)$  \cite{LM} is the infinite dimensional configuration space of a closed string. } is more suitable. In this framework, the semi-classical expansion is obtained from the geometric tubular expansion \cite{tubular} around the submanifold of vanishing loops ($\cong \cM$) in $\cLM$. The geometric data appearing in the aforementioned expansion were obtained in \cite{cut-off} by taking the $N\to \infty$ limit of the same appearing in the tubular expansion around the diagonal submanifold ($\cong \cM$) of the cut-off loop space $(\cM)^N_C$ given by the Cartesian product of $N$ copies of $\cM$ with a cyclic ordering. While NLSM is the dynamical theory with configuration space $\cLM$, the string bits NLSM or the lattice NLSM preserving all the symmetries (as constructed in this work), is the dynamical theory with configuration space $(\cM)^N_C$. 

\section{ADM decomposition}
\label{a:ADM}

Here we summarize the results of ADM decomposition of the geometry of the worldsheet $(W, \gamma)$ following \cite{kuchar}. The cylindrical worldsheet is globally hyperbolic. The map,
\bea
\xi: \bbR \times S^1 \to W ~,
\eea
is given by the diffeomorphism $\xi^a(\xih)$ where $\xih=(\tau, \s)$, ($-\infty < \tau < \infty, ~ \s \in [0,2\pi]$) with $\tau$ parametrizing timelike $\bbR$ and $\s$ the spacelike Cauchy circle $S^1$. Given the Cauchy circle $\xi^a(\tau, \s)$ at a fixed $\tau$, the corresponding timelike normal vector $n^a$ is uniquely given by,
\bea
\gamma_{ab} \del_{\s} \xi^a n^b &=& 0~, \quad \gamma_{ab} n^a n^b = -1 ~.  
\label{n-def}
\eea
We now introduce a few notations. The metric in $\xih$ system is given by,
\bea
\hat \gamma_{ab} &=& \gamma_{cd} K^c{}_a K^d{}_b ~,
\label{gamma-hat}
\eea
where, $K^a{}_b = \lt( \del \xi^a \over \del \xih^b \rt)$ is the Jacobian matrix of the transformation $\xih \to \xi$. The deformation vector,
\bea
l^a &=& \del_{\tau} \xi^a ~,
\eea
is decomposed into lapse function: $l^{\perp} = - n^a l^b \gamma_{ab}$
and shift vector $l^1 = l^a K_a{}^1$ ($K_a{}^1 = \gamma_{ab} K^b{}_1 {1\over \hat \gamma_{11} }$) in the following way,
\bea
l^a &=& n^a l^{\perp} + K^a{}_1 l^1 ~.
\eea

Given the above decomposition, the action in (\ref{action-cont}) can be written as (\ref{S-ADM}) where $\beta := l^{\perp} >0$ and $e := \sqrt{\hat \gamma_{11}} >0$ are the einbeins on $\bbR$ and $S^1$ respectively and therefore transform accordingly under $\tau$ and $\s$ reparametrization. $l:=el^1$ appearing in (\ref{gamma-hat}), on the other hand, is a covariant vector under $\tau$-reparametrization, but scalar under $\s$-reparametrization. All the $\tau$ and $\s$ derivatives in $\hat \cL$ appear with the following combinations: ${1\over \beta } \del_{\tau} $ and ${1\over e} \del_{\s}$. For example, for the non-linear sigma model (\ref{nlsm-lag}) one has,
\bea
\hat \cL &=& {1 \over 4\pi \al'} \lt[ {1\over \beta} ( \del_{\tau} - l {1\over e} \del_{\s} ) \hat Z^{\al} {1\over \beta} (\del_{\tau} - l {1\over e} \del_{\s}) \hat Z^{\beta} - {1\over e} \del_{\s} \hat Z^{\al} {1\over e} \del_{\s} \hat Z^{\beta} \rt] G_{\al \beta} (\hat Z)~. 
\label{bits-nlsm-lag-ADM}
\eea

\section{Continuum limit}
\label{a:cont-lim}

By construction, the naive continuum limit is straightforward which we discuss here. Below in \S \ref{sa:latt-arg} we first discuss how the limit applied to the lattice model discussed in \S \ref{definition} produces the right result. Then in \S \ref{sa:top-arg} we discuss the limit by viewing the lattice and the continuum as topological spaces. Both the discussions have straightforward generalization to the higher dimensional case discussed in \S \ref{generalization} .

\subsection{Lattice model argument}
\label{sa:latt-arg}

The continuum limit is given by,
\bea
h \to 0 ~, \quad N \to \infty ~, \hbox{ such that ~} hN = 2\pi ~.
\label{cont-lim}
\eea
The discrete coordinate in (\ref{sigma-p}), in the above limit, becomes a continuous parameter $\s \in [0, 2\pi]$ and,
\bea
h \sum_{p=1}^N \to \oint d\s ~.
\eea
For any discrete field $\Xh_p$ appearing in (\ref{discrete-ADM}): $\Xh_p \to \Xh(\tau, \s)$. Then the action $S_B$ in (\ref{discrete-ADM}) approaches $S$ in (\ref{S-ADM}), provided the discrete $\s$-derivative in (\ref{LDD-def}) approaches the continuum $\s$-partial-derivative. Note that according to (\ref{der-consistency}), this is necessarily true. In fact, according to (\ref{der-consistency}), this is true even at finite $h$ \footnote{At this stage, this is only an intuitive expectation which we have assumed to be true in this work. This will be established more rigorously in \cite{covlatt2, covlatt3}.}. The difference is that at finite $h$ (and $N$), there exists the continuation ambiguity which needs to be fixed consistently. In the continuum limit, this ambiguity disappears. The matching of LDD and continuum derivative in the limit (\ref{cont-lim}) can also be verified explicitly using difference calculus in the following manner.  The actions of $n$-th power of difference operators are given in (\ref{Dlt-pm-n}). On the other hand, starting from the limit definition of the first continuum derivative,
\bea
f^{(1)}(\s) \equiv \lim_{h\to 0} {1\over h} \Dlt_{\pm}[f(\s)] ~, 
\eea
(where $\Dlt_{\pm}[f(\s)] \equiv \pm f(\s+h) \mp f(\s) $) one derives for the $n$-th derivative in the continuum,
\bea
f^{(n)}(\s) = \lim_{h\to 0} {1 \over h^n} \Dlt_{\pm}^n[f(\s)] ~,
\eea
implying the following results for the discrete derivatives in (\ref{LDD-def}) in the continuum limit,
\bea
\lim_{h \to 0} \del_{\pm} [f_p] = \lim_{h \to 0} \mp{1\over h} \sum_{n=1}^{\infty} {1\over n} (\mp h)^n f^{(n)}(\s_p) \to f^{(1)}(\s_p) ~.
\eea

In the continuous covariant description in (\ref{continuous-cov}), the condition (\ref{der-consistency}) is implicitly assumed. Therefore in the continuum limit the filter should just disappear, which it does,
\bea
\cF(\xi, \{\xi_p\}) &\to & \int d^2 \xih' \beta(\xih') e(\xih') {\dlt^2 (\xi-\xi(\xih')) \over (\gamma(\xi))^{1/4} (\gamma(\xi(\xih')))^{1/4} } = 1 ~.
\eea

\subsection{Topological argument}
\label{sa:top-arg}

Our approach to construct the lattice theory has been to first consider a Cauchy foliation of the spacetime and then replace each Cauchy slice by a set of discrete points in a specific manner. The correct continuum limit requires that, as a topological space, the discrete space tend to the original Cauchy slice in the limit. Below we shall argue that this is indeed the case\footnote{The author is thankful to Indrava Roy for discussion on this.}. The topological argument is facilitated by considering an Euclidean worldsheet. However, we continue to use the same notations for the Lorentzian worldsheet introduced earlier for simplicity. Since this particular discussion is completely isolated from the rest of the paper, this does not cause any confusion. 

The discrete space under consideration is $B_1^2$ (see (\ref{B-space})), which is a closed subset of the auxiliary Euclidean worldsheet $\overline{W}^2_1$ (see footnote \ref{B-Wbar}). In ``ADM frame'' $\xih = (\tau, \s)$, $\overline{W}^2_1$ is endowed with the metric,
\bea
\hat{\gamma}_{ab} &=& \begin{pmatrix} 
\beta^2 + l^2 & e l \cr e l & e^2 
\end{pmatrix} ~,
\label{metric-euclid}
\eea 
which is obtained by analytically continuing the lapse function $\beta \to -i\beta$ in (\ref{gamma-hat-mat}). $\overline{W}^2_1$ is compact (with respect to the corresponding metric topology), hence Cauchy complete. Accordingly, $B^2_1(N)$ acquires the corresponding subspace topology. 

Let $\cH$ be the collection of all (non-empty) closed, bounded subsets of $\overline{W}^2_1$, implying, 
\bea
B^2_1(N) \in \cH ~, \quad \forall N \in \bbZ_+ ~.
\eea
$\cH$, endowed with the Hausdorff metric, is complete (see \cite{munkres}), where the Hausdorff metric is given by,
\bea
\Gamma (A, B) = \inf \{\eps | A \subset U(B, \eps) \hbox{ and } B \subset U(A, \eps) \} ~, \quad A, B \in \cH ~,
\eea 
with $U(A, \eps)$ and $U(B, \eps)$ being $\eps (>0)$-neighborhoods of $A$ and $B$ respectively. Therefore, every Cauchy sequence in $\cH$ converges to an element in $\cH$.

Given the set $\{B^2_1(N)| N\in \bbZ_+\}$, we now consider a sequence $\{B^2_1(N_n)| N_n = 2^n N~, n\in \bbZ_+ \}$, where $N>>1$ is a fixed. We shall now prove that it is a Cauchy sequence. We first calculate the Hausdorff distance between any two elements,
\bea
\Gamma( B^2_1(N_n), B^2_1(N_m)) =  \lt\{ 
\begin{array}{ll}
{1\over 2}h_{n+1} \approx {\pi \over 2^{n+1} N} & \hbox{ for } m=n+1 \cr
h_{n+1} \approx {\pi \over 2^n N} & \hbox{ for } m\geq n+2 
\end{array} \rt. ~,
\eea 
where $h_n \equiv {2\pi \over N_n}$. To arrive at the above result, we first define the $\eps$-neighborhood of $B^2_1(N)$ as,
\bea
B^2_1(N, \eps) \equiv \{ (\tau, s_p) | \tau \in [\tau_1, \tau_2], s_p \in [\s_p-\eps, \s_p+\eps], p=1,\cdots ,N \}.
\eea
Then we notice that for $m=n+1$, the minimum value of $\eps$ to be chosen in order to satisfy the conditions: $B^2_1(N_n) \subset B^2_1(N_m,\eps)$ and $ B^2_1(N_m) \subset B^2_1(N_n,\eps)$ is ${1\over 2} h_{n+1}$ and the same for $m\geq n+2$ is $h_{n+1}$.

We now assert that given any $\dlt >0$, $\exists ~n_{\dlt} \in \bbZ_+$, such that,
\bea
\Gamma(B^2_1(N_n), B^2_1(N_m)) < \dlt ~, \quad \forall n, m \geq n_{\dlt} ~.
\eea
This assertion can be easily satisfied by demanding,
\bea
{\pi \over 2^{n_{\dlt}} N} < \dlt ~, \quad \Rightarrow \quad n_{\dlt} > {\ln ({\pi \over \dlt N}) \over \ln 2} ~,
\eea
proving that the sequence under consideration is Cauchy. Since $\cH$ is complete, $\displaystyle \lim_{n\to \infty} B^2_1(N_n)$ must be within $\cH$ and therefore the limit must be $\overline{W}^2_1$ itself, which is what is required for the correct continuum limit.

\section{Leibniz rule and partial integration law}
\label{a:leibniz}

Here we discuss preliminary evidences for Leibniz rule and partial integration law given by (to reduce clutter we use the simpler notation $\del$ for LDD in this appendix),
\bea
\del(f_p g_p) &=& \del f_p g_p + f_p \del g_p ~, \label{leibniz} \\
\sum_{p=1}^N \del h_p &=& 0~,  \quad (h_{p+N} = h_p) ~, \label{partial} 
\eea
respectively. More rigorous treatments will be considered in \cite{covlatt2, covlatt3}. 

We begin with the Leibniz rule (\ref{leibniz}). This holds true for both $\del_{\pm}$ separately. Their action on a product of two functions can be written as, 
\bea
\del_{\pm} (f_p g_p) &=& \del_{\pm} f_p g_p + f_p \del_{\pm} g_p + \cA_{p \pm} ~, 
\eea
where,
\bea
\cA_{p\pm} &=& \mp {1\over h} \sum_{n\geq 1} {(\mp)^n \over n} \cA^{(n)}_{p \pm} ~, \quad \cA^{(n)}_{p \pm} = \Dlt^n_{\pm} (f_p g_p) - \Dlt_{\pm}^n f_p g_p - f_p \Dlt_{\pm}^n g_p ~,
\eea
will be referred to as ``anomaly terms''. Calculated exactly, $\cA_{p \pm}$ are expected to vanish as it is evident from the explicit results of the first few terms given below,
\bea
\cA^{(1)}_{p+} &=& {\color{violet} \Dlt_+ f_p \Dlt_+ g_p} ~, \cr
\cA^{(2)}_{p+} &=& 2 ( {\color{violet} \Dlt_+ f_p \Dlt_+ g_p}  + {\color{blue} \Dlt^2_+ f_p \Dlt_+ g_p + \Dlt_+ f_p \Dlt^2_+ g_p}) + {\color{teal}\Dlt^2_+ f_p \Dlt^2_+ g_p} ~, \cr
\cA^{(3)}_{p+} &=& 3({\color{blue} \Dlt^2_+ f_p \Dlt_+ g_p + \Dlt_+ f_p \Dlt^2_+ g_p}) + 6 {\color{teal}\Dlt^2_+ f_p \Dlt^2_+ g_p} 
+ 3 ( {\color{red} \Dlt_+^3 f_p \Dlt_+ g_p + \Dlt_+ f_p \Dlt_+^3 g_p } ) \cr
&& + 3 (\Dlt_+^3 f_p \Dlt_+^2 g_p + \Dlt_+^2 f_p \Dlt_+^3 g_p) + \Dlt_+^3 f_p \Dlt^3 g_p  ~, \cr
\cA^{(4)}_{p+} &=& 6 {\color{teal} \Dlt_+^2 f_p \Dlt_+^2 g_p} + 4 ({\color{red} \Dlt_+^3 f_p \Dlt_+ g_p + \Dlt_+ f_p \Dlt_+^3 g_p} ) \cr
&&
+ 9 (\Dlt_+^3 f_p \Dlt_+^2 g_p + \Dlt_+^2 f_p \Dlt_+^3 g_p ) + 3 (\Dlt_+^4 f_p \Dlt_+ g_p + \Dlt_+ f_p \Dlt_+^4 g_p) \cr
&& + 6 \Dlt_+^3 f_p \Dlt_+^3 g_p + 3 (\Dlt_+^4 f_p \Dlt_+^2 g_p + \Dlt_+^2 f_p \Dlt_+^4 g_p ) ~, 
\eea
where for the colored terms the cancellation is complete when substituted in $\cA_{p+}$. The terms written in black are supposed to be canceled by terms coming from $\cA^{(n)}_{p+}$ with $n=5$ and higher. A similar observation can also be made for the operator $\del_-$. The number of terms contributing to the anomaly at a given value of $n$ increases with $n$. But they all get canceled when all the infinite number of terms are taken into account. Therefore, one may conclude that Leibniz rules holds true as long as the action of $\del_{\pm}$ is well defined on each of the given lattice functions.

The partial integration law (\ref{partial}) holds true simply due to the following,
\bea
\sum_{p=1}^N \Dlt_{\pm} h_p = 0 ~, 
\eea
thanks to the periodicity condition satisfied by $h_p$. The argument becomes more subtle when $h_p = f_p g_p \cdots $ is a product of multiple functions. In this case, one may expect to establish (\ref{partial}) after using (\ref{leibniz}) which amounts to throwing away the anomaly terms first before calculating the summation in (\ref{partial}), i.e.,
\bea
\sum_{p=1}^N (\del f_g g_g \cdots + f_p \del g_p \cdots + \cdots ) &=& 0~.
\label{leibniz-partial}
\eea
We now show that it is easy to establish the above form when there are only two functions involved. In this case, eq.(\ref{leibniz-partial}) can be re-written as,
\bea
\sum_{n\geq 1} {(-1)^n \over n}  \sum_{p=1}^N t^{(n)}_p = 0~, 
\label{partial2}
\eea
where,
\bea
t^{(n)}_p &=& (\Dlt_+^n - (-1)^n \Dlt_-^n) f_p g_p + f_p (\Dlt_+^n - (-1)^n \Dlt_-^n) g_p ~.
\label{t(n)p-def}
\eea
It turns out that $\displaystyle \sum_{p=1}^N t^{(n)}_p $ vanishes for each $n \geq 1$ separately due to the following result,
\bea
\sum_{p=1}^N (f_{p+n} - f_p) &=& 0 ~, \quad n \in \bbZ ~,
\label{partial-Dlt-gen}
\eea
for any periodic function and the fact that $t_p^{(n)}$ can be written in the following form,
\bea
t^{(n)}_p &=& \sum_{r=0}^{n-1} (-1)^r \binom{n}{r} (s^{(n-r)}_{(p+n-r)} - s^{(n-r)}_p ) ~,
\label{t(n)p}
\eea
where,
\bea
s^{(n)}_p &=& f_p g_{p-n} + f_{p-n} g_p~.
\eea
We have verified the above form through explicit calculations up to $n=4$,
\bea
t^{(1)}_p &=& s^{(1)}_{p+1} - s^{(1)}_p ~,  \cr
t^{(2)}_p &=& s^{(2)}_{p+2} - s^{(2)}_p - 2 (s^{(1)}_{p+1} - s^{(1)}_p) ~,  \cr
t^{(3)}_p &=& s^{(3)}_{p+3} - s^{(3)}_p - 3(s^{(2)}_{p+2} - s^{(2)}_p) + 3 (s^{(1)}_{p+1} - s^{(1)}_p) ~, \cr
t^{(4)}_p &=& s^{(4)}_{p+4} - s^{(4)}_p - 4(s^{(3)}_{p+3} - s^{(3)}_p) + 6(s^{(2)}_{p+2} - s^{(2)}_p) - 4(s^{(1)}_{p+1} - s^{(1)}_p)   ~.
\eea
Therefore, assuming that the form (\ref{t(n)p}) is correct, (\ref{partial2}) is satisfied because for each $n$ the summation over $p$ vanishes. 

For more than two fields in each additive factor in (\ref{leibniz-partial}), the structure of the term $t^{(n)}_p$ in (\ref{t(n)p}) is no more valid and the above argument does not work. At this stage, we do not have such a direct method of establishing (\ref{partial2}) in this case. The question will be dealt with using a different approach in \cite{covlatt2, covlatt3} which will fully establish (\ref{leibniz}) and (\ref{partial}). 

\section{Some technicalities}
\label{a:technical}

\subsection{Consistency between continuum and discrete calculations}
\label{sa:consistency}

In \S\ref{s:cov} we discussed two approaches to study the lattice/bits theory covariantly, namely continuous and discrete covariant approaches. Here we shall demonstrate that consistency of these calculations relies upon the following identity,
\bea
h \sum_{p'=1}^N \dlt(\s_p-\s_{p'}) f(\s_{p'}) = f_p = \sum_{p'=1}^N \dlt_{p, p'} f_{p'} ~.
\label{consistency}
\eea
This can be proved in a straightforward manner by substituting,
\bea
f(\s) &=& \sum_{n\in \bbZ} \tilde f_n e^{in \s} ~, \cr
\dlt (\s-\s') &=& {1\over 2\pi} \sum_{n\in \bbZ} e^{in(\s-\s')} ~,
\eea
on the left hand side and recognizing that, 
\bea
\dlt_{m, n} &=& {1\over N} \sum_{p=1}^N e^{i(m-n)p{2\pi \over N} } ~.
\eea
Notice that $\{f_p\}$ has only a finite number of degrees of freedom and so is true for its continuation $f(\s)$ and Fourier coefficients $\{\tilde f_n\}$, most of the elements of the latter being zero. Nonetheless, the Fourier transform written above is valid. 

To demonstrate the relevance of (\ref{consistency}) we perform a specific set of calculations namely, the ones that lead to the results in (\ref{action-var-bits}). This also serves the purpose of establishing the local correspondence in (\ref{local-correspondence}). We begin by noting how the calculation goes in the continuum. Given (\ref{action-cont}) and (\ref{action-var-cont}), one arrives at,
\bea
\cE(\cX(\xi)) &=& \int d^2\xi' {\dlt \over \dlt X(\xi)} \lt\{ \sqrt{\gamma (\xi')} \cL (\cX(\xi')) \rt\} ~.
\eea
The integrand on the right hand side has the following structure,
\bea
\sqrt{\gamma(\xi') } \sum_{n\geq 0} (-1)^n \cC^{a_1\cdots a_n}_n(\cX(\xi')) D'_{a_1} \cdots D'_{a_n} \dlt^2(\xi', \xi) ~, 
\eea
where we have incorporated the situation where the action contains arbitrary higher derivative terms and used the first equation of (\ref{fund-func-derv}). After performing partial integrations, one finds,
\bea
\cE(\cX(\xi)) &=& \cC_0 (\cX(\xi)) + D_a \cC_1^a(\cX(\xi) + D_b D_a \cC_2^{ab}(\cX(\xi)) + \cdots  ~.
\label{cE-cont}
\eea
The bits version of the above analysis in the continuum approach will be given as follows,
\bea
\dlt_X S_B &=& \int d^2 \xi \sqrt{\gamma (\xi)} \dlt X(\xi) \cF(\xi, \{\xi_p\}) \int d^2 \xi' \lt[ {\dlt \over \dlt X(\xi)} 
\lt\{ \sqrt{\gamma (\xi')} \cL(\cX(\xi')) \rt\} \rt] \cF(\xi', \{\xi_{p'}\})  ~, \cr
&&
\eea
where we have used field independence of the filter. The functional partial derivative kept in the square brackets produces the same result as in the continuum discussed above. The subsequent partial integrations remain unaffected by the filter as the latter is covariantly constant. This leads to the following result,
\bea
\dlt_X S_B &=& \int d^2 \xi \sqrt{\gamma (\xi)} \dlt X(\xi) \cE(\cX(\xi)) \cF(\xi, \{\xi_p\})  \cF(\xi, \{\xi_{p'}\})  ~.
\eea
One of the filters can be integrated out by going to the ADM frame. This gives,
\bea
\dlt_X S_B &=& h^2 \sum_{p,p'=1}^N \int d\tau (\hat \gamma(\hat \xi_p) \hat \gamma(\hat \xi_{p'}))^{1/4} \dlt(\s_p-\s_{p'}) 
\dlt X(\xi_p) \cE(\cX(\xi_p)) ~.
\eea
One now uses the left part of (\ref{consistency}) to arrive directly at the second line of (\ref{action-var-bits}). 

In the discrete covariant approach one needs to perform the calculation in ADM frame,
\bea
\dlt_X S_B &=& h \sum_{p=1}^N \int d\tau \sqrt{\hat \gamma_p (\tau)} \dlt \hat X_p(\tau)  h \sum_{p'=1}^N \int d\tau' {\dlt \over \dlt \hat X_p (\tau)} 
\lt\{ \sqrt{\hat \gamma_{p'}(\tau')} \hat \cL(\hat \cX_{p'} (\tau')) \rt\} ~. \cr
&&
\eea
In this case the structure of the second integrand takes the following form,
\bea
{1\over h} \sqrt{\hat \gamma_{p'}(\tau')} \sum_{n\geq 0} (-1)^n \hat \cC^{a_1 \cdots a_n}_n(\hat \cX_{p'}(\tau')) \hat D'_{a_1} \cdots \hat D'_{a_n} \dlt^2_{(p',p)}(\tau', \tau) ~,
\eea
where we have used the second equation of (\ref{fund-func-derv}). Then using the discrete version of the partial integration law (see Appendix \ref{a:leibniz}) one arrives at the second line of (\ref{action-var-bits}) with, 
\bea
\cE(\cX_p(\tau)) &=& \cC_0 (\cX_p(\tau)) + D_a \cC_1^a(\cX_p(\tau)) + D_b D_a \cC_2^{ab}(\cX_p(\tau)) + \cdots   
~,
\label{cE-disc}
\eea
which is the lattice analogue of (\ref{cE-cont}).

\subsection{Gauss's law}
\label{sa:gauss}

The discrete version (\ref{gauss-discrete}) can be established simply by writing the integrand on the LHS in ADM frame and using the discrete covariant derivative in that frame,
\bea
LHS &=& \sum_{p=1}^N \int_{\tau_1}^{\tau_2} d\tau \hat \del_a \lt[ \sqrt{\hat \gamma_p(\tau)} \hat J_p^a (\tau) \rt] ~.
\eea
For $a=\s$ the contribution vanishes due to partial integration law (Appendix \ref{a:leibniz}), leaving us with the result as given in (\ref{gauss-discrete}). 

To establish the continuum covariant form in (\ref{gauss-cont-cov}), one proves,
\bea
\int_{\Wb^2_1} d^2\xi \sqrt{\gamma(\xi)} D_a J^a (\xi) \cF(\xi, \{\xi_p\}) &=& h \sum_{p=1}^N \int_{\tau_1}^{\tau_2} d\tau \sqrt{\hat \gamma_p(\tau)} D_a J_p^a ~, \label{gauss-cont-cov-1} \\
\int_{\del \Wb^2_1} dx \sqrt{\gammalb (x)} n_a(x) J^a (\xi(x)) \cF(\xi(x), \{\xi_p\}) &=& h \sum_{p=1}^N \lt[ \sqrt{\hat \gamma_p(\tau)} \hat J^0_p(\tau) \rt]^{\tau_2}_{\tau_1} ~. \label{gauss-cont-cov-2} 
\eea
To prove the first one, one notices that the LHS can alternatively be written as,
\bea
LHS_{(\ref{gauss-cont-cov-1})} &=& \int_{\overline W} d^2 \xi \sqrt{\gamma(\xi)} D_a J^a (\xi) \cF_1^2(\xi, \{\xi_p\}) ~, \cr
&=& \int_{\overline W} d^2 \xi \sqrt{\gamma(\xi)} \lt[ D_a \{J^a (\xi) \cF_1^2(\xi, \{\xi_p\}) \} - J^a(\xi) D_a \cF_1^2 (\xi, \{\xi_p\}) \rt] ~,
\label{gauss-cont-cov-1-lhs} \cr
&&
\eea
where $\overline W$ is an infinite auxiliary worldsheet, while,
\bea
\cF_1^2(\xi, \{\xi_p\}) &=& h \sum_{p=1}^N \int_{\tau_1}^{\tau_2} d\tau \sqrt{\hat \gamma (\hat \xi_p) } \dlt^2(\xi, \xi_p) ~,
\label{filter-cov-finite}
\eea
is a finite filter as opposed to (\ref{filter-cov}) which has infinite range for the $\tau$-integration. The first term in (\ref{gauss-cont-cov-1-lhs}) is a total derivative term on $\overline W$. Therefore it vanishes, as the boundary of $\overline W$ does not have a support for the finite filter. For the second term, we use (\ref{my-id-bits}) and the bits version of (\ref{cov-const-cP}) to find,
\bea
LHS_{(\ref{gauss-cont-cov-1})} &=& h \sum_{p=1}^N \int_{\tau_1}^{\tau_2} d\tau \sqrt{\hat \gamma(\hat \xi_p)}  D_{p b}  \int_{\overline W} d^2 \xi \sqrt{\gamma(\xi)} J^a(\xi) \cP_a{}^b (\xi, \xi_p) \dlt^2 (\xi, \xi_p) ~, \cr
&&
\eea
which is same as the RHS of (\ref{gauss-cont-cov-1}). 

To prove (\ref{gauss-cont-cov-2}), we first note that the ADM frame constructed in Appendix \ref{a:ADM} is adapted to a specific, but arbitrary, Cauchy slicing. In the present calculation, given the two boundary slices $C_1$ and $C_2$, we first consider a Cauchy slicing of the entire finite cylinder with initial and final slices given by $C_1$ and $C_2$ respectively and then choose the ADM frame to be adapted to that slicing. Therefore, for any given slice $C$, the internal coordinate $x$ is transformable to $\s$ at the corresponding value of $\tau$ implying,
\bea
x \to \s ~, \quad \sqrt{\underline{\gamma}} \to e(\tau, \s) ~, \quad n_a \to \begin{pmatrix}
\pm \beta(\tau, \s) \cr 0
\end{pmatrix} ~,
\eea
where the sign in the last expression needs to be chosen following the discussion below eq.(\ref{gauss-cont-cov}). Therefore, we could write, for the LHS of (\ref{gauss-cont-cov-2}),
\bea
LHS_{(\ref{gauss-cont-cov-2})} &=& \lt[ \int d\s e(\tau, \s) \beta(\tau, \s)  \hat J^{\tau}(\tau, \s) \cF (\hat \xi(\s), \{\hat \xi_p\}) \rt]^{\tau_2}_{\tau_1} ~, \cr
&=& h \lt[ \sum_{p=1}^N \int_{\bbR} d\tau' \sqrt{\hat \gamma_p(\tau')} \int d\s e(\tau, \s) \beta(\tau, \s) \hat J^{\tau}(\tau, \s) 
{\dlt(\tau-\tau') \dlt(\s-\s_p) \over (\hat \gamma_p(\tau') \hat \gamma (\tau, \s))^{1/4} } \rt]^{\tau_2}_{\tau_1} ~, \cr
&=& h \sum_{p=1}^N \lt[ \sqrt{\hat \gamma_p(\tau)} \hat J_p^{\tau}(\tau) \rt]^{\tau_2}_{\tau_1} ~,
\eea
which is the right hand side of (\ref{gauss-cont-cov-2}). 

\section{Configuration space approach to symmetries}
\label{a:symmetry}

Here we analyze symmetries of the bits theory in the configuration space approach following the formulation of \cite{holten}. We generalize the criteria for global and local symmetries to a generic theory with diffeomorphism invariance. After developing the general formalism, we specialize to string bits NLSM. 

\subsection{General framework}
\label{sa:general}

We consider the general action for the bits as given in (\ref{continuous-cov}). The condition that an infinitesimal variation is a symmetry is given by,
\bea
\dlt^{symm} S_B &=& \int_{\overline W_1^2} d^2\xi \sqrt{\gamma (\xi)} D_a \rmj^a (\cX(\xi)) \cF(\xi, \{\xi_p\}) ~, \cr
&=& \int_{\del \overline W^2_1} dx \sqrt{\underline{\gamma} (x)} n_a (x) \rmj^a (\cX(\xi)) \cF(\xi(x), \{\xi_p \})  ~,
\label{symm-def-bits}
\eea
where we have taken the $\tau$ integral in $S_B$ to be within the range $[\tau_1, \tau_2]$ and used Gauss's theorem (\ref{gauss-cont-cov}) to get to the second line. Given this we define the following,
\bea
B(\{C_p\}) &\equiv & \int_C dx \sqrt{\underline{\gamma}(x)} n_a(x) \rmj^a(\cX(\xi(x))) \cF(\xi(x), \{\xi_p \}) ~,
\label{B-def-bits}
\eea
where $C$ is any Cauchy circle and $n^a$ is the corresponding timelike, future directed unit normal. $C_p$ is the intersection between $C$ and the $p$-th bit-worldline given by $\xi_p$. Therefore,
\bea
\dlt^{symm} S_B &=& B(\{{C_2}_p\})-B(\{{C_1}_p\}) ~.
\eea

To proceed further and define the conserved charge, let us specialize to the following case,
\bea
\cX &=& \{ \gamma_{ab}, Z^{\mu}, \del_a Z^{\mu} \}~,
\label{field-content}
\eea
of which NLSM (\ref{nlsm-lag}) is an example. A straightforward manipulation reveals that an arbitrary variation of the action can be written as,
\bea
\dlt S_B &=& \int_{\overline W_1^2} d^2 \xi \sqrt{\gamma (\xi)} \lt[ \cE(\gamma)^{ab} \dlt \gamma_{ab} + \cE_{(Z) \mu} \dlt Z^{\mu} + D_a \lt({\del \cL \over \del (\del_a Z^{\mu}) } \dlt Z^{\mu} \rt)\rt]_{\xi} \cF(\xi, \{\xi_p\}) ~, 
\label{bits-action-arbit-var}
\cr
&&
\eea
where, the equations of motion (in the continuum) for $\gamma_{ab}$ and $Z^{\mu}$ are given by,
\bea
\cE_{(\gamma)}^{ab} &\equiv & {\del \cL \over \del \gamma_{ab}} + {1\over 2} \gamma^{ab} \cL = 0 ~, \cr
\cE_{(Z) \mu} &\equiv & {\del \cL \over \del Z^{\mu}} - D_a \lt( {\del \cL \over \del (\del_a Z^{\mu}) }\rt) = 0 ~,
\label{bits-EOM-gen}
\eea
Note that the same set of equations, with subscript $p$ added, holds true for the bits. The latter is obtained simply by integrating the filter out in (\ref{bits-action-arbit-var}). This is a an example of the local correspondence discussed in \S \ref{laws}. Therefore, on a ``classical path'' (as indicated by the subscript $cl$ below),
\bea
\dlt_{cl} S_B &=& \int_{\overline W^2_1} d^2 \xi \sqrt{\gamma_{cl}(\xi)} D_a \lt[ \cP_{cl \mu}^a (\dlt Z^{\mu})_{cl} \rt]_{\xi} \cF_{cl}(\xi, \{\xi_p\}) =: 
A_{cl}(\{{C_2}_p\}) - A_{cl}(\{{C_1}_p\}) ~, \label{A-def-bits-1} \cr
&&
\eea
where, 
\bea
\cP^a_{\mu} = {\del \cL \over \del (\del_a Z^{\mu}) } ~, \quad A(\{C_p\}) &=& \int_C dx \sqrt{\underline{\gamma} } n_a(x) \cP^a_{\mu} \dlt Z^{\mu}|_{\xi(x)} \cF(\xi(x), \{\xi_p\}) ~,
\label{A-def-bits-2}
\eea
and the ``classical filter'' $\cF_{cl}(\xi, \{\xi_p\})$ is given by (\ref{filter-cov}) with $\hat \gamma_{ab}$ and $\gamma_{ab}$ replaced by their on-shell values\footnote{Note however that in 2D the metric is always diffeomorphic to a conformally flat form and that the NLSM we shall eventually consider is insensitive to the conformal factor.}. Therefore, for a symmetry variation around a classical path, 
\bea
\dlt^{symm}_{cl} S_B &=& B_{cl}(\{{C_2}_p\}) - B_{cl}(\{{C_1}_p\}) = A_{cl}(\{{C_2}_p\}) - A_{cl}(\{{C_1}_p\}) ~,
\eea
and the corresponding classical conserved charge is given by,
\bea
G(\{C_p\}) &=& A_{cl}(\{C_p\}) - B_{cl}(\{C_p\}) = \int_C dx \sqrt{\underline{\gamma_{cl}} (x)} n_a(x) J^a_{cl}|_{\xi(x)} 
\cF_{cl}(\xi(x), \{\xi_p \}) ~, \label{J-def-bits-1} 
\label{G-def-bits} \cr
&&
\eea
where the total current is given by,
\bea
J^a &\equiv & \cP^a_{\mu} \dlt Z^{\mu} - {\mathrm j}^a ~.
\label{J-def-bits-2} 
\eea

A local conservation equation is formulated in the following way. Consider two infinitesimally closed Cauchy slices $C$ and $C'$ such that $C'$ occurs later than $C$ and compute,
\bea
d G(\{C_p\}) &=& G(\{C'_p\})-G(\{C_p\}) ~, \cr
&=& \int_{\cV} d^2 \xi \sqrt{\gamma_{cl}(\xi)} D_a J_{cl}^a(\cX(\xi))  \cF_{cl}(\xi, \{\xi_p \}) ~, \cr
&=& 0 ~,
\label{dG(C)}
\eea
where $\cV$ is the infinitesimal cylinder bounded by $C$ and $C'$. Going along the same line of argument as given in the proof of (\ref{gauss-cont-cov-2}), we consider an ADM frame that fits to the boundary slices $C$ and $C'$. Computing in this frame, we may write,
\bea
{d G (\tau) \over d\tau } &=& \sum_{p=1}^N \sqrt{\hat \gamma_{cl_p} } D_a J_{cl}^a(\cX(\xi_p)) = 0 ~,
\eea
where we have replaced the argument $\{C_p\}$ by $\tau$ as in the ADM frame Cauchy slices occur at fixed values of $\tau$. Notice that the above result is true for an arbitrary classical solution and an arbitrary Cauchy slice. Moreover, according to our definition of $J$ (see eqs. (\ref{symm-def-bits}, \ref{A-def-bits-2}, \ref{J-def-bits-2}), the latter contains arbitrary variation parameters. Therefore, one concludes that the following ``local equation'' must hold everywhere on the lattice,
\bea
D_aJ^a_{cl} (\cX_p(\tau)) &=& 0 ~.
\eea
This is another important example of the local correspondence discussed earlier.

\subsection{Classification of global and local symmetries}
\label{sa:classification}

While the previous subsection formulates the condition of a symmetry transformation and explicitly constructs the corresponding conserved charge, our discussion needs to be extended to distinguish local and global symmetries. Given the local correspondence as demonstrated in the previous section, it is enough to perform the analysis in the continuum only. Therefore the problem at hand is to generalize Holten's analysis in \cite{holten} to the case of a generally covariant theory with field content (\ref{field-content}). 

We consider symmetry transformations generated by the following set of infinitesimal fields\footnote{Eventually, these deformations will correspond to Weyl, diffeomorphism and spacetime isometries respectively of the NLSM.},
\bea
\{\omg, \kappa^a, V^{\mu}_i \} ~, \quad i =1,\cdots , I~,
\eea
where $\omg$ and $V^{\mu}_i$ are worldsheet scalar and $\kappa^a$ is a worldsheet vector. $V^{\mu}_i$ is a target space vector for each $i$. The infinitesimal transformations of the fields, namely $\dlt \gamma_{ab}$ and $\dlt Z^{\mu}$, the current $\rmj^a$ introduced in (\ref{symm-def-bits}) and the total current $J^a$ in (\ref{J-def-bits-2}), admit covariant derivative expansion of the following form, 
\bea
\dlt \gamma_{ab} &=& \sum_{n\geq 0} \lt[R_n^{(\gamma; \omg)}{}_{ab}{}^{c_1\cdots c_n} D_{c_1\cdots c_n} \omg 
+ R_n^{(\gamma; \kappa)}{}_{ab}{}^{c c_1 \cdots c_n} D_{c_1 \cdots c_n} \kappa_c 
+ \sum_i R_n^{(\gamma; V_i)}{}_{ab}{}_{\mu}{}^{c_1 \cdots c_n} D_{c_1 \cdots c_n} V_i^{\mu} \rt] ~, \cr
\dlt Z^{\mu} &=& \sum_{n\geq 0} \lt[R_n^{(Z; \omg) \mu c_1\cdots c_n} D_{c_1\cdots c_n} \omg 
+ R_n^{(Z; \kappa)\mu c c_1 \cdots c_n} D_{c_1 \cdots c_n} \kappa_c 
+ \sum_i R_n^{(Z; V_i)\mu}{}_{\nu}{}^{c_1 \cdots c_n} D_{c_1 \cdots c_n} V_i^{\nu} \rt] ~, \cr
\rmj^a &=& \sum_{n\geq 0} \lt[R_n^{(\rmj; \omg)a c_1\cdots c_n} D_{c_1\cdots c_n} \omg 
+ R_n^{(\rmj; \kappa) a c c_1 \cdots c_n} D_{c_1 \cdots c_n} \kappa_c 
+ \sum_i R_n^{(\rmj; V_i) a}{}_{\mu}{}^{c_1 \cdots c_n} D_{c_1 \cdots c_n} V_i^{\mu} \rt] ~. \cr
J^a &=& \sum_{n\geq 0} \lt[G_n^{(\omg)a c_1\cdots c_n} D_{c_1\cdots c_n} \omg 
+ G_n^{(\kappa) a c c_1 \cdots c_n} D_{c_1 \cdots c_n} \kappa_c 
+ \sum_i G_n^{(V_i) a}{}_{\mu}{}^{c_1 \cdots c_n} D_{c_1 \cdots c_n} V_i^{\mu} \rt] ~, \cr
&&
\label{der-expansions}
\eea
where we have used the notation: $D_{c_1 \cdots c_n} = D_{c_1} \cdots D_{c_n}$. The worldsheet derivatives of $V$ should be calculated in the following manner,
\bea
D_c V^{\mu} &=& D_c Z^{\nu} \nabla_{\nu} V^{\mu} ~, \cr
D_{c_1 c_2} V^{\mu} &=& D_{c_1} (D_{c_2} Z^{\nu} \nabla_{\nu} V^{\mu}) 
= D_{c_1} D_{c_2} Z^{\nu} \nabla_{\nu} V^{\mu} + D_{c_2} Z^{\nu_2} D_{c_1} Z^{\nu_1} \nabla_{\nu_1} \nabla_{\nu_2} V^{\mu} ~, \cr
&&
\eea
and so on. The coefficients in (\ref{der-expansions}) should satisfy relations according to (\ref{J-def-bits-2}),
\bea
G_n^{(\omg)a c_1\cdots c_n} &=& \cP^a_{\mu} R_n^{(Z; \omg) \mu c_1\cdots c_n} - R_n^{(\rmj; \omg)a c_1\cdots c_n} ~, \cr
G_n^{(\kappa)a cc_1\cdots c_n} &=& \cP^a_{\mu} R_n^{(Z; \kappa) \mu c c_1\cdots c_n} - R_n^{(\rmj; \kappa)a c c_1\cdots c_n} ~, \cr
G_n^{(V_i)a}{}_{\nu}{}^{c_1\cdots c_n} &=& \cP^a_{\mu} R_n^{(Z; V_i) \mu}{}_{\nu}{}^{c_1\cdots c_n} - R_n^{(\rmj; V_i) a}{}_{\nu}{}^{c_1\cdots c_n} ~.
\label{Gn-Rn}
\eea
Given this, the expansion for the conserved charge (i.e. the continuum version of (\ref{G-def-bits}) is given by,
\bea
G(C) 
&=& \int_C dx \sqrt{\underline{\gamma_{cl}} (x)} n_a \sum_{n\geq 0} \lt[ G_n^{(\omg) ac_1 \cdots c_n} D_{c_1\cdots c_n} \omg 
+ G_n^{(\kappa) abc_1\cdots c_n} D_{c_1 \cdots c_n} \kappa_b \rt. \cr
&& \lt. + \sum_i G_n^{(V_i) a}{}_{\mu}{}^{c_1\cdots c_n}  D_{c_1 \cdots c_n} V^{\mu}_i \rt]_{cl} ~.
\eea
Computing in the adapted ADM frame, one further finds for the time derivative of the charge,
\bea
{dG\over d\tau} 
&=& \int d\s \sqrt{\hat \gamma_{cl}} \lt[ D_a G_0^{(\omg) a} \omg + D_a G_0^{(\kappa) ab} \kappa_b + \sum_i D_a G_0^{(V_i) a}{}_{\mu} V^{\mu}_i \rt. \cr
&&
+ \sum_{n\geq 0} \lt\{ (D_a G_{n+1}^{(\omg) ac_1 \cdots c_{n+1}} + G_n^{(\omg) c_1 \cdots c_{n+1}} ) D_{c_1\cdots c_{n+1} } \omg \rt. \cr
&& 
+ (D_a G_{n+1}^{(\kappa) abc_1\cdots c_{n+1}} + G_n^{(\kappa) c_1bc_2\cdots c_n} ) D_{c_1 \cdots c_{n+1}} \kappa_b  \cr
&& \lt. \lt. + \sum_i \lt(D_a G_{n+1}^{(V_i) a}{}_{\mu}{}^{c_1\cdots c_{n+1}} + G_n^{(V_i) c_1}{}_{\mu}{}^{c_2\cdots c_{n+1} } \rt) D_{c_1 \cdots c_{n+1} } V^{\mu}_i \rt\} \rt]_{cl} ~, \cr
&=& 0 ~.
\eea
Since the deformation fields and all their higher derivatives are independent, we must have,
\bea
D_a G_0^{(\omg) a} = 0 ~, && D_a G_{n+1}^{(\omg) ac_1 \cdots c_{n+1}} + G_n^{(J; \omg) c_1 \cdots c_{n+1}} = 0 ~, \cr
D_a G_0^{(\kappa) ab} = 0~, && D_a G_{n+1}^{(\kappa) abc_1\cdots c_{n+1}} + G_n^{(\kappa) c_1bc_2\cdots c_n} = 0 ~, \cr
D_a G_0^{(V_i) a}{}_{\mu} = 0 ~, && D_a G_{n+1}^{(V_i) a}{}_{\mu}{}^{c_1\cdots c_{n+1}} + G_n^{(V_i) c_1}{}_{\mu}{}^{c_2\cdots c_{n+1} } = 0 ~.
\label{Gn-constraints}
\eea

We now define a symmetry to be global if all the contributions coming from higher derivatives of the deformation vanish in ${dG\over d\tau}$ leading to the conservation laws given by the equations on the left column above involving $G_0$'s. In this case, the conserved change is given by,
\bea
G(C) &=& \int_C dx \sqrt{\underline{\gamma_{cl}} (x)} n_a \lt[ G_0^{(\omg) a} \omg + G_0^{(\kappa)ab} \kappa_b + \sum_i G_0^{(V_i)a}{}_{\mu} V_i^{\mu}   \rt]~.
\label{global-charge-gen-cont}
\eea
Any symmetry not satisfying the above condition is a local one. We now repeat the same argument from \cite{holten}. In a situation (which is more standard) where only a finite number of higher derivative terms are involved in the expansions in (\ref{der-expansions}), there exists an integer $m$, such that $G_n = 0$, $\forall n\geq m$, where $G_n$ is any coefficient appearing on the left hand sides of (\ref{Gn-Rn}). Then, by the equations on the right column of (\ref{Gn-constraints}), we must have $G_{m-1} = 0$. The argument then continues to all the lower values of $n$ leading to the fact that $G_n=0~, \forall n\geq 0$. Therefore, for a local symmetry the conserved charge vanishes which gives rise to a constraint that remains the same under time evolution.


\subsection{Application to string bits NLSM}
\label{sa:app-nlsm}

We begin by considering isometries of the target space and show that these are global symmetries of the bits NLSM. The transformations are given by (\ref{nlsm-global}). The variation of the action in the continuum is\footnote{Although we are studying the bits theory, we often write equations that refer to the continuum. We do this to reduce clutter and whenever it does not cause any confusion, given the local correspondence.}, 
\bea
\dlt^{symm} S \propto \sum_i \eta_i \int d^2 \xi \sqrt{\gamma} \gamma^{ab} \del_a Z^{\mu} \del_b Z^{\nu} (\nabla_{\mu} V_{i\nu} + \nabla_{\nu} V_{i\mu})~,
\eea
vanishes due to (\ref{killing}), implying that $\rmj^a$ vanishes. This, along with: $\cP^a_{\mu} \propto \gamma^{ab} G_{\mu \nu}(Z) \del_b Z^{\nu}$ imply that the only non-vanishing $G_n$ in (\ref{Gn-Rn}) is,
\bea
G_0^{(V_i) a}{}_{\mu} \propto \eta_i \gamma^{ab} G_{\mu \nu} \del_b Z^{\nu} ~.
\eea
This leads to the global charge (\ref{global-charge}), where we have used the notation $Q$ instead of $G$. The conservation equation is given by,
\bea
{d Q\over d\tau} &=& \int d\s \sqrt{\hat \gamma_{cl}} \lt[ D_b G_0^{(V_i)a}{}_{\mu} V_i^{\mu} + G_0^{(V_i)a}{}_{\mu} D_b V_i^{\mu} \rt]_{cl} = 0 ~,
\eea
where the two terms vanish separately. The first term vanishes due to the equation of motion,
\bea
\gamma_{cl}^{ab} \lt[ \del_a \del_b Z_{cl}^{\mu} - \kappa^c_{cl ab} \del_c Z_{cl}^{\mu} + \Gamma^{\mu}_{\nu \rho} \del_a Z_{cl}^{\nu} \del_b Z_{cl}^{\rho} \rt] = 0 ~,
\eea
$\kappa^c_{ab}$ being the worldsheet Christoffel symbols, while the second term vanishes due to (\ref{killing}). The latter verifies our criterion for global symmetry that all terms involving derivatives of the deformation field must vanish. 

The local symmetries are given by (\ref{nlsm-local}). From the variation of the action one finds,
\bea
\rmj^a &=& \kappa^a \gamma^{bc} G_{\mu \nu} \del_b Z^{\mu} \del_c Z^{\nu} ~,
\eea
implying the following results,
\bea
G_n^{(\omg)ac_1 \cdots c_n} &=& 0~, \quad \forall n \geq 0 ~, \\
G_0^{(\kappa)ab} &\propto& T^{ab} ~, \quad G_n^{(\kappa)a c c_1\cdots c_n} = 0 ~, \quad \forall n \geq 1 ~,
\eea
where the stress tensor is given by (\ref{stress-tensor}), leading to the local charge (\ref{local-charge}). One can now calculate,
\bea
{dG\over d\tau} = \int d\s \sqrt{\hat \gamma_{cl}} \lt[ D_a G_0^{(\kappa)a b} \kappa_b + G_0^{(\kappa) ab} D_a \kappa_b  \rt]_{cl} ~.
\eea
We notice that there are finite number of derivative terms involving the deformation field. Hence according to our criterion, these are local symmetries, implying the constraints (\ref{constraints}). 

\section{Delta function identities}
\label{a:delta-ids}

In different calculations we encounter dealing with derivatives of delta functions both in the continuum (see e.g. \S \ref{sa:gauss}) and on the lattice (see e.g. (\ref{der-kronecker-1}), (\ref{der-kronecker-2})). The goal of this appendix is to present a unified treatment for the same. 

We begin with the following standard identities in the continuum (see e.g. \cite{dewitt1}),
\bea
(\xi-\xi')^{a_1} (\xi-\xi')^{a_2} \cdots (\xi-\xi')^{a_n} \dlt^D (\xi, \xi') &=& 0 ~, \quad \hbox{for } n\geq 1~, \cr 
(\xi-\xi')^{a_1} (\xi-\xi')^{a_2} \cdots (\xi-\xi')^{a_n} \del_b \dlt^D (\xi, \xi') &=& - \dlt_{n, 1} \dlt^{a_1}{}_b \dlt^D (\xi, \xi') ~, \cr
\del_a \dlt^D(\xi, \xi') &=& - \del_{a'} \dlt (\xi, \xi') - \kappa^b_{b a} \dlt^D(\xi, \xi') ~.
\label{cont-dlt-ids}
\eea
where the subscripts $a$ and $a'$ to the derivative symbol refer to the coordinates $\xi$ and $\xi'$ respectively. $\kappa^a_{bc}$ are the Christoffel symbols computed from the metric $\gamma_{ab}$. For the last identity, we would like to use a more covariant (though equivalent) version given by,
\bea
D_{a} \dlt^D(\xi, \xi') &=& - \cP_{a}{}^{a'}(\xi, \xi') D_{a'} \dlt^D(\xi, \xi') ~,
\label{cov-id}
\eea
where 
\bea
\cP_{a}{}^{a'}(\xi, \xi') &=& \tilde e_{(b) a}(\xi) e^{(b) a'}(\xi') ~,
\eea 
is the bi-vector of geodesic parallel transport \cite{dewitt2}. $e^{(a)}{}_b$ are the vielbeins with the index kept in round brackets being the internal index. $\tilde e^{(a)}{}_b(\xi)$ represents $e^{(a)}{}_b(\xi')$ parallel transported to $\xi$ along the geodesic. By Taylor expanding $\tilde e^{(a)}{}_b(\xi)$ around $\xi'$ and using the first two identities in (\ref{cont-dlt-ids}), one shows that (\ref{cov-id}) is same as the last one in (\ref{cont-dlt-ids}). 

It is also possible to establish (\ref{cov-id}) by proving the following result,
\bea
&& \int d^D \xi \sqrt{\gamma(\xi)} \int d^D \xi' \sqrt{\gamma(\xi')} D_a\dlt^D(\xi, \xi') X^a(\xi) Y(\xi') \cr
&=& - \int d^D \xi \sqrt{\gamma(\xi)} \int d^D \xi' \sqrt{\gamma(\xi')} \cP_a{}^{a'}(\xi, \xi') D_{a'}\dlt^D(\xi, \xi') X^a(\xi) Y(\xi') ~,
\label{cov-id-proof}
\eea
where $X$ and $Y$ are arbitrary vector-valued and scalar test fields respectively both of which vanish at the boundary. This is a straightforward exercise once the following identity is used,
\bea
D_{a'} \cP_{a}{}^{a'}(\xi, \xi') \dlt^D(\xi, \xi') &=& 0 ~,
\label{cov-const-cP}
\eea
which can in turn be established by again Taylor expanding $\cP_{a'}{}^a(\xi, \xi')$ around $\xi=\xi'$ and using the first result in (\ref{cont-dlt-ids}).

The entire discussion can now be repeated for the bits. In particular, the first two equations in (\ref{cont-dlt-ids}) hold true with $\xi$ replaced by $\xi_p$. For the first equation, this claim is trivially true. To establish the bits version of the second equation (with the derivative appearing on the LHS being LDD), one simply follows through the discrete analogue of the continuum argument by using the Leibniz rule and partial integration law for LDD as discussed in Appendix \ref{a:leibniz}. We now consider the bits version of (\ref{cov-id}), 
\bea
D_{p b} \dlt^2(\xi_p, \xi) &=& - \cP_b{}^a(\xi_p, \xi) D_a \dlt^2(\xi_p, \xi) ~. 
\label{my-id-bits}
\eea
One way to established this is to first consider the continuum version of the left hand side,
\bea
D_{p b} \dlt^2(\xi_p, \xi) &=& \int d^2 \xi' \sqrt{\gamma (\xi')} \dlt^2 (\xi_p, \xi') \cP_b{}^{c'}(\xi_p, \xi') D_{c'} \dlt^2(\xi', \xi) ~,
\eea
and then use (\ref{cov-id}),
\bea
D_{p b} \dlt^2(\xi_p, \xi) &=& - \int d^2 \xi' \sqrt{\gamma (\xi')} \dlt^2 (\xi_p, \xi') \cP_b{}^{c'}(\xi_p, \xi') \cP_{c'}{}^c (\xi', \xi) D_c \dlt^2(\xi', \xi) ~, \cr
&=& - \cP_b{}^a(\xi_p, \xi) D_a \dlt^2 (\xi_p, \xi) ~.
\eea
Alternatively, one could prove the bits versions of (\ref{cov-id-proof}),
\bea
&& \sum_p \int d\tau \sqrt{\hat \gamma(\hat \xi_p)} \int d^D \xi' \sqrt{\gamma(\xi')} D_{pa} \dlt^D(\xi_p, \xi') X_p^a Y(\xi') \cr
&=& - \sum_p \int d\tau \sqrt{\hat \gamma(\hat \xi_p)} \int d^D \xi' \sqrt{\gamma(\xi')} \cP_a{}^{a'}(\xi_p, \xi') D_{a'}\dlt^D(\xi_p, \xi') X_p^a Y(\xi') ~,
\label{cov-id-proof-bits}
\eea
which is as easy to prove as (\ref{cov-id-proof}) once the Leibniz rule and partial integration law of Appendix \ref{a:leibniz} are used. 

Finally, we consider the identities (\ref{der-kronecker-1}, \ref{der-kronecker-2}). 

\vspace{.1in}
\noindent
{\bf Proof of (\ref{der-kronecker-1}):} It is straightforward to prove this directly.  Using the results,
\bea
\Dlt_+^n f_p = (-1)^n \sum_{r=0}^n (-1)^r \binom{n}{r} f_{p+r} ~, && \Dlt_-^n f_p = \sum_{r=0}^n (-1)^r \binom{n}{r} f_{p-r} ~,
\label{Dlt-pm-n}
\eea
one first shows,
\bea
\Dlt_{\pm}^n \dlt_{p, p'} &=& (-1)^n {\Dlt'_{\mp} }^n \dlt_{p, p'} ~.
\eea
This is then used further to show,
\bea
\del_{\pm} \dlt_{p, p'} &=& - \del'_{\mp} \dlt_{p, p'} ~,
\eea
leading to (\ref{der-kronecker-1}). 

\vspace{.1in}
\noindent
{\bf Proof of (\ref{der-kronecker-2}):} Just like the continuum analogue, this result can also be proved by performing lattice Taylor expansion\footnote{This will be discussed in detail in \cite{covlatt2}.} of $f_{p'}$ around $p$ and then using the lattice analogues of the first two equations in (\ref{cont-dlt-ids}). Another approach is to show,
\bea
\sum_{p, p'} g_p f_{p'} \del \dlt_{p, p'} &=& \sum_{p, p'} \lt[ g_p f_p \del \dlt_{p, p'} + g_p \del f_p \dlt_{p, p'} \rt] = \sum_p g_p \del f_p  ~,
\eea
for an arbitrary periodic test field $g_p$. In the second step, we have taken the summation over $p'$ inside the derivative for the first term. The above result can easily be established by the direct method like in the previous case or use the properties of Appendix \ref{a:leibniz}. This indicates that although the Kronecker delta can be viewed as a discrete function, LDD acting on it behaves more like a distribution.

\begin{center}
{\bf Acknowledgement}
\end{center}

The author is thankful to Ramesh Anishetty, A. P. Balachandran, Sumanto Chanda, Sumit R. Das, Ghanashyam Date, Rajiv V. Gavai, Debashis Ghoshal, Jaya N. Iyer, M. Padmanath, Chandrima Paul, Indrava Roy, Ashoke Sen, Alfred D. Shapere and Sayantan Sharma for useful discussions over years. The author is also grateful to Sumit R. Das and Alfred D. Shapere for their continuous encouragement. Talks based on this work were presented in {\it HRI Workshop of String Theory: Developments in String Perturbation Theory}, February 1-13, 2016 Harish-Chandra Research Institute, Allahabad, India; {\it Chennai Strings Meeting}, November 21-23, 2019 The Institute of Mathematical Sciences, Chennai, India; {\it Theory Seminar}, September 2020, Department of Physics and Astronomy, University of Kentucky, USA; {\it Institute Seminar Days}, November 2020, The Institute of Mathematical Sciences, Chennai, India. The author has benefited from the discussions that followed.

\end{document}